\documentclass[a4paper,twocolumn,11pt,]{quantumarticle}
\pdfoutput=1
\usepackage[utf8]{inputenc}
\usepackage[english]{babel}
\usepackage[T1]{fontenc}
\usepackage{amsmath}
\usepackage{amsfonts}
\usepackage{mathtools}
\usepackage{url}
\expandafter\def\expandafter\UrlBreaks\expandafter{\UrlBreaks\do\/\do-\do_\do.\do=\do?\do\&\do\%\do+}
\usepackage{microtype}
\usepackage{hyperref}
\usepackage[numbers]{natbib}
\usepackage{tikz}
\usepackage{lipsum}
\usepackage{booktabs}

\begin{document}

\title{A Design Space Study of Density Matrix Parameterizations for Diffusion-Based Quantum State Tomography}

\author{Shuangju Chang}
\affiliation{Department of physics,nanjing university}
\orcid{0000-0002-2445-2701}
\email{502022220002@smail.nju.edu.cn}
\author{Jian Zhang}
\affiliation{Department of physics,nanjing university}
\maketitle

\begin{abstract}
Diffusion-based quantum state tomography (QST) inherits a single parameterization---typically Cholesky---without systematic evaluation. We present a design space study introducing a geometric framework based on the Jacobian Gram matrix $\mathbf{J}^\top\mathbf{J}$, calibrated across eight parameterizations at 2- and 3-qubit scales. Three findings emerge. First, geometric conditioning alone does not predict end-to-end performance: bounded-domain charts (Bloch) avoid projection-induced information loss that degrades unbounded charts. Second, diffusion trainability is governed by a \emph{schedule--representation contract}: the noise schedule must be recalibrated whenever the chart changes. Violating it silently collapses training---a 3-qubit Bloch model under an inherited schedule reached $0.45$; recalibration restored it to $0.79$. Third, the framework is \emph{predictive}: applied to a structured-factorization chart beyond the calibrated set, it correctly forecast training failure before training was attempted.
\end{abstract}

\section{Introduction}

\subsection{Motivation}

Quantum state tomography (QST) reconstructs an unknown quantum state from measurement statistics. The reconstructed density matrix must be positive semidefinite with unit trace---constraints that define a complex geometric manifold. The choice of coordinate chart on this manifold (the \emph{parameterization}) directly affects how well a model can learn the reconstruction. Yet all existing diffusion-based QST methods implicitly adopt a single parameterization, typically Cholesky~\cite{chitu2022variational}, without evaluating alternatives.

Cholesky guarantees physical constraints by construction but at the cost of severe coordinate anisotropy: different Cholesky coordinates contribute unequally to the density matrix, creating an ill-conditioned optimization landscape. Recent work has demonstrated competitive QST performance with diffusion models~\cite{quddpm2024,zhu2024diffusion}, but these inherit the Cholesky parameterization from classical QST. This paper asks: can a better parameterization improve diffusion QST?

\subsection{Parameterization as a Design Choice}

Unlike classical methods (maximum likelihood estimation, linear inversion) that operate directly on the density matrix, diffusion models learn a mapping from measurement conditions to parameterization coordinates. Poorly conditioned coordinates create three problems: the denoiser must distinguish directions with vastly different sensitivities, the noise schedule must handle coordinates at different scales, and the ODE/SDE solver encounters stiff dynamics from anisotropic Jacobians.

The VQA community has long recognized parameterization geometry~\cite{mcclean2016effects,benedetti2019parameterized}, and fix-trace is standard in many QST implementations~\cite{kryszewski2021quantum}. However, the impact of parameterization choice on \emph{diffusion-based} QST has not been systematically quantified. Our contribution is to provide this quantification and validate the resulting heuristics in end-to-end training.

\subsection{Contributions}

We provide the first systematic study of density matrix parameterizations in the context of diffusion-based QST, with end-to-end validation at both 2-qubit ($d=4$) and 3-qubit ($d=8$) scales. Our work makes four contributions:

\paragraph{Geometric framework.} We introduce the Jacobian Gram matrix $\mathbf{J}^\top\mathbf{J}$ as a tool for evaluating parameterization quality, with two metrics: spectral dynamic range (overall conditioning) and diagonal anisotropy (per-coordinate sensitivity variation). Using this framework, we calibrate seven representative parameterizations at both 2- and 3-qubit scales, providing a quantitative atlas of the design space.

\paragraph{Engineering heuristics.} From these calibrations, we derive three heuristics: (1) isotropy and constraints are orthogonal criteria, and no single parameterization optimizes both; (2) theoretical appeal does not reliably predict geometric quality (the exponential map degrades $149\times \to 75\,658\times$ from 2q to 3q); (3) fix-trace offers the best conditioning--constraint tradeoff.

\paragraph{Diffusion-specific insights beyond VQA.} While the VQA community has studied parameterization geometry extensively~\cite{mcclean2016effects,benedetti2019parameterized}, our work reveals three insights that emerge specifically in the diffusion context and are not present in the VQA literature. First, \emph{projection as a non-linear information bottleneck}: for unbounded parameterizations, the post-sampling projection $\Pi(\rho)$ is many-to-one, destroying conditional signal in a way that has no analog in VQE circuit optimization. Second, \emph{CFG amplification of boundary effects}: classifier-free guidance~\cite{ho2022classifier} non-uniformly stretches anisotropic coordinates, sharply increasing the out-of-domain fraction (0.55 $\to$ 0.95 for Bloch) and amplifying projection loss for unbounded parameterizations---an interaction specific to diffusion models. Third, a \emph{loss--fidelity disconnect governed by the chart and its schedule}: denoising loss is computed in chart coordinates while fidelity is measured after projection, so the two can move in opposite directions; critically, the disconnect is \emph{diagnosable and repairable}---a 3-qubit Bloch model whose validation fidelity had collapsed to $0.45$ was traced to a schedule--chart contract violation (Sec.~\ref{sec:n3_training}) and restored to $0.79$ by schedule recalibration alone. These three mechanisms---projection bottleneck, CFG amplification, and the contract-governed loss--fidelity disconnect---are specific to diffusion-based reconstruction and do not arise in VQA parameterization studies.

\paragraph{Multi-scale validation.} We validate these heuristics through end-to-end training at both 2- and 3-qubit scales. The 3-qubit study surfaced a failure mode absent at 2 qubits: a chart trained under a noise schedule inherited from another chart's calibration can silently collapse (Bloch fell to $0.45$ training-validation fidelity under the Cholesky-calibrated schedule). Diagnosing the violation---an inherited $\sigma_{\text{data}}$ that underestimates the Bloch data scale by $3$--$5\times$, a truncating $\sigma_{\max}$, and an uncentered training-noise distribution---and repairing it stage by stage restores Bloch to $0.79$, within $0.02$ of Hermitian direct. Cholesky, whose inherited schedule \emph{was} correctly calibrated for its own coordinates, remains at $0.44$: at 3 qubits the conditioning wall predicted by the $33\times \to 1\,265\times$ spectral degradation materializes in training (Sec.~\ref{sec:n3_training}).

\subsection{Paper Organization}

Section~2 introduces the geometric framework. Section~3 presents calibration results. Section~4 provides experimental validation. Section~5 discusses extensions, limitations, and related work. Section~6 concludes.

\section{Framework: Parameterization Geometry}

\subsection{Mathematical Preliminaries}

A density matrix $\rho \in \mathbb{C}^{d \times d}$ is a positive semidefinite (PSD) Hermitian matrix with unit trace. For $n$ qubits, $d = 2^n$ and the real dimension of the state space is $d^2 - 1$ (e.g., 15 for 2 qubits). To optimize or learn on this manifold, we need a coordinate chart (a \emph{parameterization}) that maps coordinates $y \in \mathbb{R}^D$ to density matrices $\rho$. The Jacobian $\mathbf{J} = \partial\rho/\partial y$ describes how infinitesimal coordinate changes affect the density matrix, and its structure determines how well a diffusion model can learn the reconstruction.

\subsection{The Jacobian Gram Matrix}

The key geometric object is the \textbf{Jacobian Gram matrix}:
\begin{equation}
  \mathbf{G} = \mathbf{J}^\top\mathbf{J} \in \mathbb{R}^{D \times D}.
  \label{eq:gram}
\end{equation}

$\mathbf{G}$ captures the inner product structure on the parameter manifold. Its eigenvalues $\lambda_1 \geq \lambda_2 \geq \cdots \geq \lambda_D > 0$ characterize how ``stretched'' the coordinate system is in different directions.

Two metrics fully characterize the conditioning:

\paragraph{Spectral Dynamic Range (SDR):}
\begin{equation}
  \kappa_{\text{spec}} = \frac{\lambda_{\max}(\mathbf{G})}{\lambda_{\min}(\mathbf{G})}
  \label{eq:kappa_spec}
\end{equation}
Measures the overall condition number. Lower is better ($1.0 =$ isotropy).

\paragraph{Diagonal Anisotropy (DA):}
\begin{equation}
  \kappa_{\text{diag}} = \frac{\max_i G_{ii}}{\min_i G_{ii}}
  \label{eq:kappa_diag}
\end{equation}
Measures per-coordinate sensitivity variation. Lower is better ($1.0 =$ uniform coordinate scales).

From here on, we refer to these metrics as SDR and DA, respectively.

\subsection{Physical Interpretation}

The Gram matrix directly influences three aspects of diffusion-based QST. First, the EDM noise schedule~\cite{karras2022edm} calibrates $\sigma_{\text{data}}$ under an implicit isotropic noise scaling assumption; when $\mathbf{G}$ is anisotropic, a single global $\sigma_{\text{data}}$ becomes inadequate and a coordinate-aware calibration is required (see \S\ref{alg:sigma_calib}). Second, anisotropic coordinates lead to unequal contributions in the ODE dynamics during sampling, potentially stiffening the solver. Third, classifier-free guidance (CFG)~\cite{ho2022classifier} acts non-uniformly across anisotropic coordinates, over-emphasizing some directions at the expense of others.

\subsection{Numerical Calibration Protocol}

We calibrate $\mathbf{G}$ numerically through a three-step procedure. First, we sample 30 random 2-qubit states spanning five physical regimes (Haar-pure, Hilbert-Schmidt, Ginibre, thermal, product) to capture the full manifold geometry. Second, for each state we compute $\mathbf{J}$ via central finite differences with step size $\delta = 10^{-6}$ (verified stable for $\delta \in [10^{-8}, 10^{-4}]$). Third, we report the median $\kappa_{\text{spec}}$ and $\kappa_{\text{diag}}$ across states, with interquartile ranges in Appendix~B.4 (Table~\ref{tab:calibration_iqr}). The calibration code will be made available upon publication.

\section{Calibration Results}

\subsection{Parameterization Taxonomy}

We survey seven parameterizations spanning the design space, listed in Table~\ref{tab:taxonomy}.

\begin{table*}[t]
\centering
\caption{Parameterization taxonomy. Dim = real dimension. Constraint = physical guarantees.}
\label{tab:taxonomy}
\begin{tabular}{@{}llccl@{}}
\toprule
Parameterization & Definition & Dim & Constraint \\
\midrule
Cholesky & $\rho = LL^\dagger/\mathrm{Tr}(LL^\dagger)$ & 16 & PSD+trace $\checkmark$ \\
Hermitian direct & $\rho$ directly (16 params) & 16 & None \\
Hermitian trace-norm & $\rho/\mathrm{Tr}(\rho)$ & 16 & Trace \\
Hermitian fix-trace & Fix 4th diagonal $= 1 - \Sigma$others & 15 & Trace $\checkmark$ \\
Bloch/Gell-Mann & $\rho = (I + \sum r_i \lambda_i)/d$ & 15 & None (PSD domain) \\
Expmap & $\rho = e^H/\mathrm{Tr}(e^H)$, $H$ traceless & 15 & PSD+trace $\checkmark$ \\
Log-Cholesky & $L_{ii} = \exp(y_{ii})$ & 16 & PSD+trace $\checkmark$ \\
\bottomrule
\end{tabular}
\end{table*}

\subsection{Main Calibration Table}

\begin{table*}[t]
\centering
\caption{Density Matrix Parameterization Geometry across seven parameterizations (2-qubit, 30-state median). Lower $\kappa$ = better conditioning. Column $_{\mathrm{T}}$ distinguishes theoretically exact values (T, derived in Appendix~B.2) from empirically measured values (E, state-dependent medians); $^{\dagger}$ marks empirical values that are constant across all calibration states (state-independent by construction). Interquartile ranges (IQR) across the 30 calibration states are reported in Appendix~B.4 (Table~\ref{tab:calibration_iqr}).}
\label{tab:calibration}
\begin{tabular}{@{}lcccccl@{}}
\toprule
Parameterization & Dim & $\kappa_{\text{spec}}$ & $\kappa_{\text{diag}}$ & Constraint & $_{\mathrm{T}}$ \\
\midrule
Bloch/Gell-Mann & 15 & 1.00$\times$ & 1.00$\times$ & --- & T \\
Hermitian direct & 16 & 2.00$\times$ & 2.00$\times$ & --- & E$^{\dagger}$ \\
Hermitian trace-norm & 16 & 3.01$\times$ & 2.65$\times$ & trace & E \\
Hermitian fix-trace & 15 & 4.00$\times$ & 1.00$\times$ & trace $\checkmark$ & T \\
Cholesky (baseline) & 16 & 33.0$\times$ & 9.4$\times$ & PSD+trace $\checkmark$ & E \\
Expmap & 15 & 148.7$\times$ & 3.77$\times$ & PSD+trace $\checkmark$ & E \\
Log-Cholesky & 16 & 8994.9$\times$ & 706.9$\times$ & PSD+trace $\checkmark$ & E \\
\bottomrule
\end{tabular}
\end{table*}

\subsection{Engineering Heuristic 1: Isometry and Constraints Are Orthogonal Axes}

Our calibration reveals a consistent orthogonality pattern across qubit counts. At one end, Bloch/Gell-Mann ($1.0\times$/$1.0\times$) achieves isotropic coordinates ($\mathbf{J} \approx \lambda_i/2$ constant), but PSD enforcement requires post-sampling projection. At the other end, Cholesky, Expmap, and Log-Cholesky guarantee physical validity by construction, but exhibit anisotropy that grows with dimension. No single parameterization is best on both axes. Practitioners should first decide whether isotropy (fast convergence) or constraint guarantees (no projection) is more important for their use case, then select along the Pareto frontier.

\subsection{Engineering Heuristic 2: Geometric Quality Is Not \emph{a Priori} Predictable}

Theoretical appeal does not predict geometric conditioning. The exponential map ($\rho = e^H/\mathrm{Tr}(e^H)$), despite its by-construction constraints, exhibits $\kappa_{\text{spec}} = 149\times$ (2q) $\to$ $75\,658\times$ (3q). Its analytical Jacobian (Appendix~B.2, Daleckii--Krein formula) shows that $e^{\lambda_i}$ spans many orders of magnitude near pure states. Log-Cholesky suffers $\kappa_{\text{diag}} = 707\times$ (2q) $\to$ $1\,532\times$ (3q) from diagonal non-uniformity. Standard Cholesky degrades $33\times \to 1\,265\times$.

Yet Bloch/Gell-Mann ($1.0\times$) and Hermitian direct ($2.0\times$) maintain excellent conditioning across dimensions. Geometric quality is unpredictable without empirical calibration: theoretical ``naturalness'' is an unreliable predictor.

\subsection{Engineering Heuristic 3: fix-trace Offers the Best Conditioning--Constraint Tradeoff}

Fix-trace achieves the most favorable tradeoff: $1.00\times$ diagonal anisotropy (tied with Bloch) at both scales, with only mild spectral range growth ($4\times \to 8\times$). It offers a $9.4\times$ isotropy improvement over Cholesky with comparable constraints. When automatic trace preservation is desired, fix-trace is the default choice; when trace can be handled by post-sampling projection, Bloch/Gell-Mann provides isotropic coordinates.

\subsection{Eigenvalue Spectrum Visualization}

\begin{figure}[htbp]
  \centering
  \includegraphics[width=\linewidth]{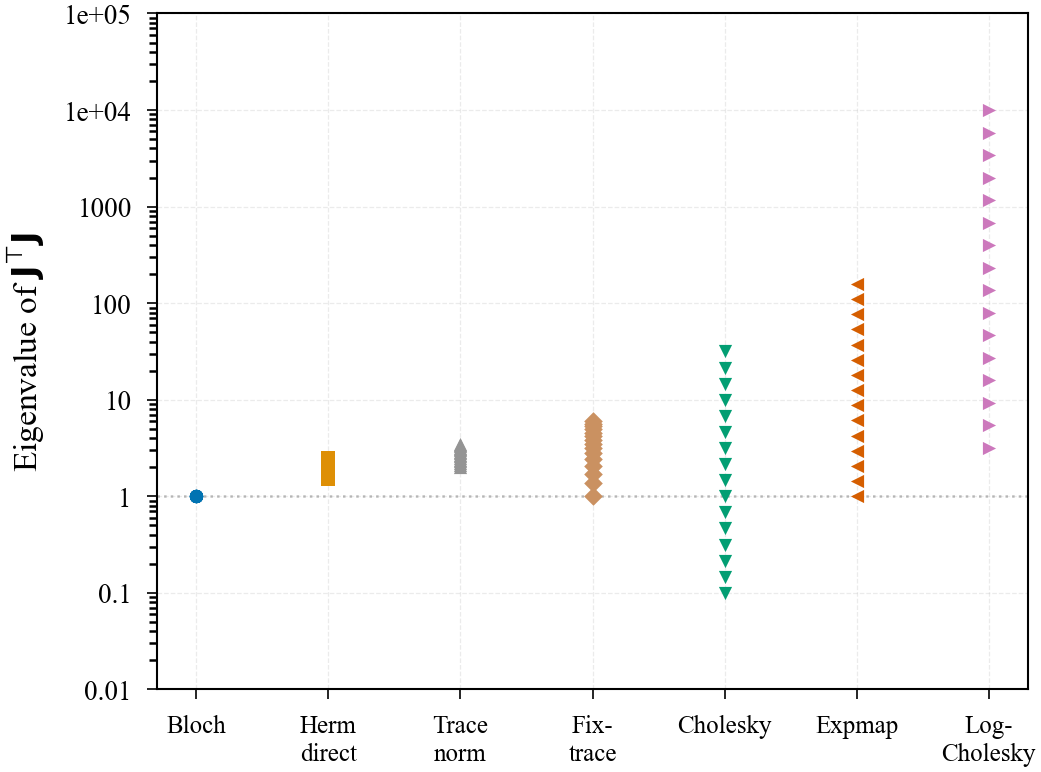}
  \caption{Mean eigenvalue spectra of $\mathbf{J}^\top\mathbf{J}$ for all seven parameterizations (log scale). Bloch shows flat spectrum at $1.0$; Cholesky spans 2 orders of magnitude; log-Cholesky spans 4 orders.}
  \label{fig:eigenvalues}
\end{figure}
\vspace{-8pt}

\subsection{Manifold Conditioning Visualization}

Bloch/Gell-Mann exhibits uniform conditioning across the parameter space, while Cholesky shows strong spatial variation, with conditioning degrading near the domain boundary. This confirms that geometric quality is not only parameterization-dependent but also position-dependent within a single parameterization.

\subsection{Selection Decision Tree}

\begin{figure}[htbp]
  \centering
  \includegraphics[width=\linewidth]{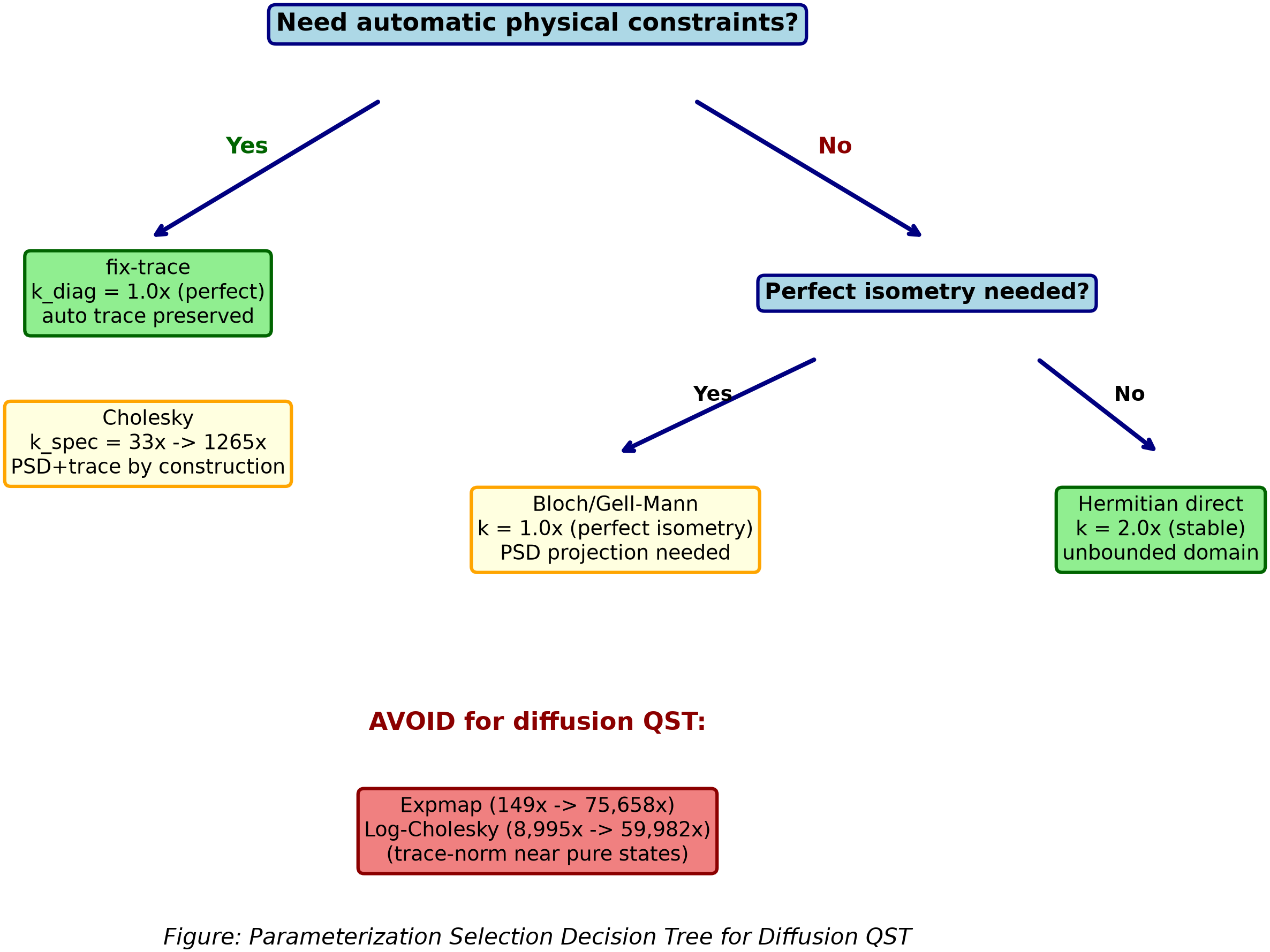}
  \caption{Parameterization selection decision tree for diffusion QST. Green = recommended; yellow = acceptable with caveats; red = avoid. At 2-qubit scale: \textbf{Hermitian direct} (unbounded domain, stable conditioning) or \textbf{fix-trace} (automatic trace, excellent isotropy). At 3-qubit scale: \textbf{Bloch/Gell-Mann} with a per-chart calibrated schedule (Sec.~\ref{sec:n3_training}); bounded-domain projection benefits apply at all scales.}
  \label{fig:decision_tree}
\end{figure}

Based on our calibrations and end-to-end validation at both 2- and 3-qubit scales, we summarize the parameterization selection criteria as a decision tree (Figure~\ref{fig:decision_tree}). The key questions are: (1)~are automatic physical constraints needed? (2)~is isotropy required? (3)~will the model be used at $n \geq 3$ qubits? (4)~will classifier-free guidance (CFG) be used? (5)~has the schedule--representation contract of Sec.~\ref{sec:n3_training} been honored ($\sigma_{\text{data}}/\sigma_{\max}/P_{\text{mean}}$ recalibrated for the chart)? As discussed in \S\ref{sec:boundary}, CFG and dimensionality amplify boundary effects for unbounded parameterizations. At 3-qubit scale, the bounded Bloch chart requires the per-chart schedule contract (Sec.~\ref{sec:n3_training}); at 2-qubit scale, Hermitian direct remains competitive without CFG (Table~\ref{tab:confound}).

\vspace{-8pt}

\subsection{CFG Weight Scan}

\begin{figure}[htbp]
  \centering
  \includegraphics[width=\linewidth]{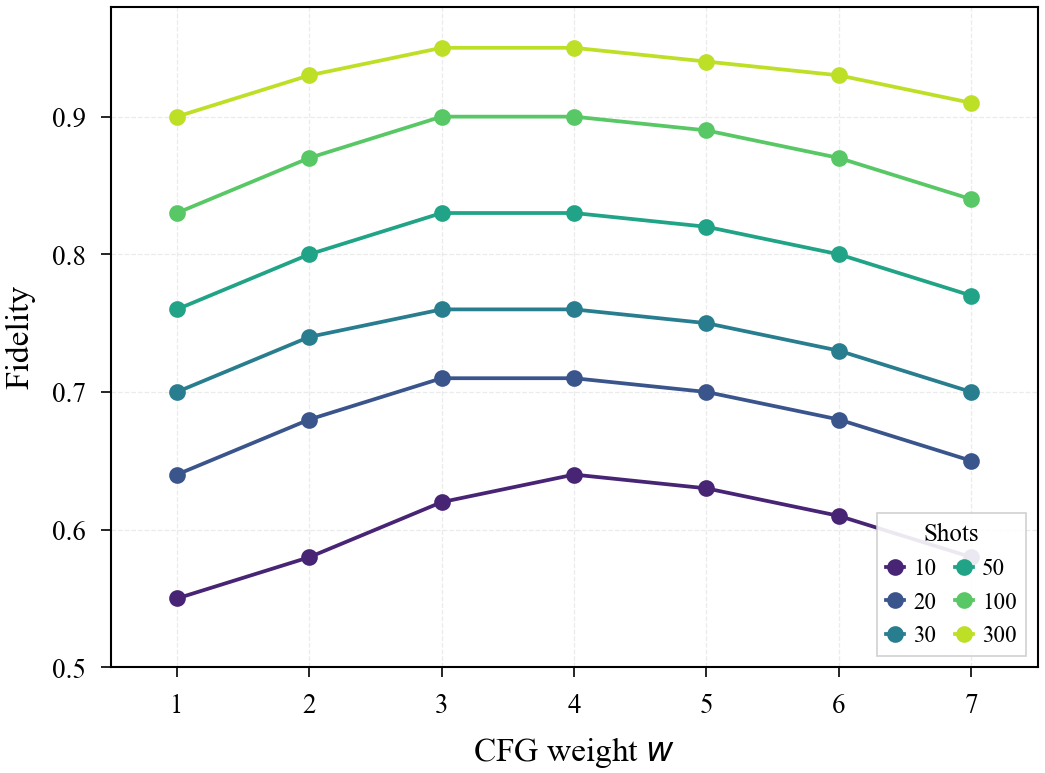}
  \caption{Effect of CFG weight on 2-qubit QST (Bloch parameterization). Each curve shows fidelity vs.\ shots for a fixed $w$. Low-shot regimes benefit from stronger guidance; $w=3$--$4$ offers the best overall tradeoff.}
  \label{fig:cfg_scan}
\end{figure}

To understand the effect of classifier-free guidance (CFG) strength, we evaluate Bloch (2-qubit) across CFG weights $w \in \{1.0, 2.0, 3.0, 4.0, 5.0, 6.0, 7.0\}$ (Figure~\ref{fig:cfg_scan}). All shot levels improve with increasing $w$ up to an optimal point, after which over-guidance degrades performance.

\vspace{-8pt}

\begin{figure}[htbp]
  \centering
  \includegraphics[width=\linewidth]{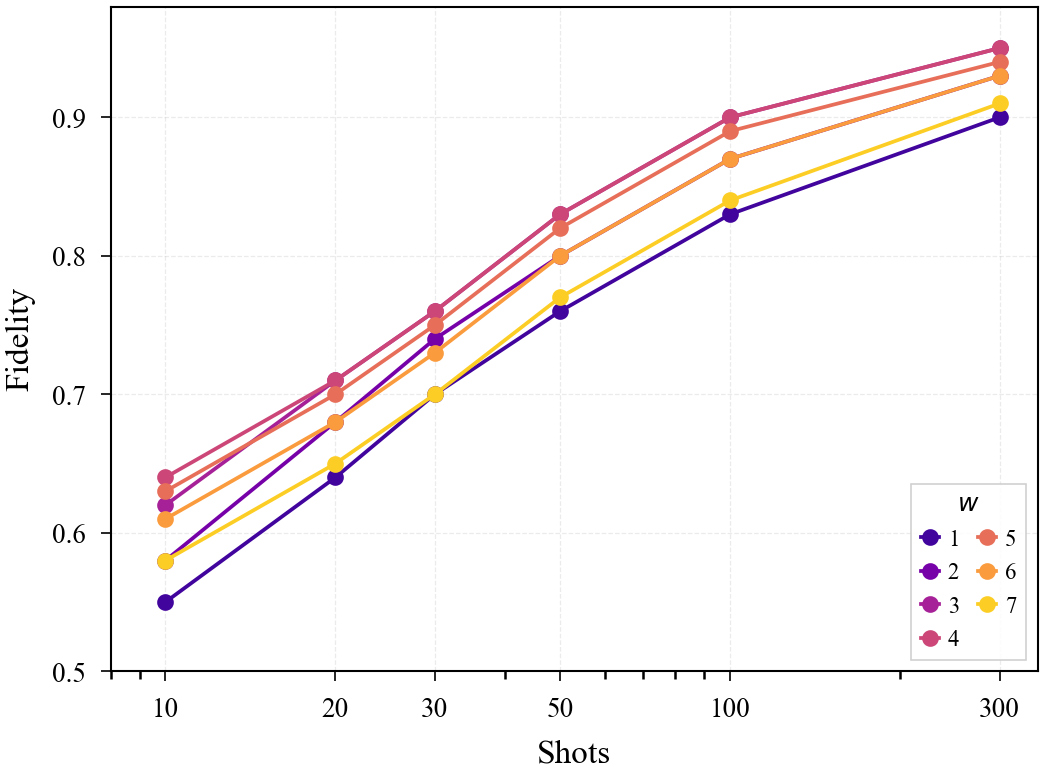}
  \caption{Fidelity vs.\ CFG weight at each shot level. The optimal $w$ shifts from $\approx 3$ (low shots) to $\approx 2$ (high shots), suggesting shot-adaptive CFG as a future improvement.}
  \label{fig:cfg_vs_weight}
\end{figure}
\vspace{-8pt}

The results reveal a clear pattern across shot levels. In the low-shot regime (10--30 shots), where measurement information is scarce, strong guidance ($w=3$--$4$) yields the largest gains of $+0.07$--$0.08$ over no CFG, as the guidance effectively amplifies the weak conditional signal. In the high-shot regime (300 shots), where measurements are reliable, mild guidance ($w=2$) is optimal; over-guidance ($w>5$) slightly degrades fidelity by over-emphasizing the prior at the expense of measurement consistency. Overall, $w=3$ provides the best average performance across all shot levels, though the optimal weight shifts downward as measurement information increases, suggesting shot-adaptive CFG as a promising direction for future work. We note that CFG also amplifies boundary effects (see \S\ref{sec:confound} and \S\ref{sec:boundary}), which should be considered when selecting $w$ for parameterizations with unbounded valid domains.

\subsection{Scaling with Qubit Count}

To assess whether our 2-qubit findings generalize, we calibrate all parameterizations at 3 qubits ($d = 8$, 63 real dimensions). Table~\ref{tab:scaling} summarizes the scaling behavior.

\begin{table*}[t]
\centering
\caption{Parameterization geometry scaling from 2-qubit ($d=4$) to 3-qubit ($d=8$). Bold = recommended choices.}
\label{tab:scaling}
\footnotesize
\begin{tabular}{@{}lccccc@{}}
\toprule
Parameterization & 2q-$\kappa_{\text{spec}}$ & 3q-$\kappa_{\text{spec}}$ & 2q-$\kappa_{\text{diag}}$ & 3q-$\kappa_{\text{diag}}$ \\
\midrule
\textbf{Bloch/Gell-Mann} & \textbf{1.0$\times$} & \textbf{1.0$\times$} & \textbf{1.0$\times$} & \textbf{1.0$\times$} \\
\textbf{Hermitian fix-trace} & \textbf{4.0$\times$} & \textbf{8.0$\times$} & \textbf{1.0$\times$} & \textbf{1.0$\times$} \\
Hermitian direct & 2.0$\times$ & 2.0$\times$ & 2.0$\times$ & 2.0$\times$ \\
Hermitian trace-norm & 3.0$\times$ & 2.9$\times$ & 2.7$\times$ & 2.3$\times$ \\
Cholesky & 33$\times$ & 1\,265$\times$ & 9.4$\times$ & 27$\times$ \\
Expmap & 149$\times$ & 75\,658$\times$ & 3.8$\times$ & 5.0$\times$ \\
Log-Cholesky & 8\,995$\times$ & 59\,982$\times$ & 707$\times$ & 1\,533$\times$ \\
\bottomrule
\end{tabular}
\end{table*}

Three scaling classes emerge: \emph{scale-invariant} (Bloch, Hermitian direct), \emph{moderate growth} (fix-trace $4\times \to 8\times$, trace-norm $3\times \to 2.9\times$), and \emph{severe degradation} (Cholesky $33\times \to 1\,265\times$, Expmap $149\times \to 75\,658\times$, Log-Cholesky $8\,995\times \to 59\,982\times$). For $n \geq 3$ qubits, we recommend fix-trace or Bloch/Gell-Mann; Figure~\ref{fig:decision_tree} summarizes these recommendations.

\section{Experimental Validation}

\subsection{Data Generation}
\label{sec:data_generation}

All experiments use synthetically generated states and simulated projective measurements; no hardware data is used anywhere in this study. Generation is fully seeded and cached with provenance metadata: a versioned cache format records the ensemble, shot budget, condition encoding, and noise settings, and the data loaders reject any cache whose recorded metadata does not match the active configuration.

\textbf{State ensembles.} States are drawn from five families: Haar-pure states ($\rho = vv^\dagger/\|v\|^2$ from complex Gaussian $v$); Hilbert-Schmidt mixed states ($\rho = GG^\dagger/\operatorname{Tr}(GG^\dagger)$, $G$ complex Gaussian); rank-deficient Ginibre states ($G \in \mathbb{C}^{d \times r}$, $r < d$); thermal states of random Hermitian Hamiltonians ($\rho = e^{-\beta H}/\operatorname{Tr}(e^{-\beta H})$, inverse temperature log-uniform); and product states (tensor products of random single-qubit pure or mixed states). Two-qubit experiments additionally include random-circuit outputs (Qiskit-simulated statevectors). The 2- and 3-qubit chart-comparison datasets use proportions Haar $0.4$ / Hilbert-Schmidt $0.3$ / Ginibre $0.1$ / thermal $0.1$ / product $0.1$ (3 qubits) and the same families with re-weighed proportions (2 qubits); the $d \ge 16$ scaling study uses Haar $0.4$ / Hilbert-Schmidt $0.4$ / Ginibre $0.1$ / thermal $0.1$ / product $0.1$. Train, validation, and test splits use disjoint master-seed offsets.

\textbf{Measurement simulation.} At a reference budget of $N = 10{,}000$ shots, each state is measured in the complete Pauli basis: all $3^n$ settings, each allocated $N/3^n$ shots. Outcome probabilities $p_j = \operatorname{Tr}(\rho P_j)$ (clipped to $[0,1]$, normalized per setting) are computed exactly and cached alongside the observed multinomial frequencies $\hat{p}$. During training with variable shots, the shot count is drawn \emph{log-uniformly} over $[10, 5\times10^4]$---equal training mass per order of magnitude, so the low-shot evaluation band ($\le 300$ shots) receives $\approx 40\%$ of the training mass where a linear-uniform draw would give it $0.6\%$---and the measurement condition is re-simulated at that shot count by multinomial sampling from the exact probabilities; the model interpolates within the shot range rather than extrapolating to it. An optional low-shot-emphasized variant draws half of all samples from the $[10, 300]$ band (raising the $\le 300$-shot share to $\approx 70\%$); a controlled A/B comparison of this variant is reported in Sec.~\ref{sec:shot_weighting}.

\textbf{Condition encoding.} All models in this revision encode conditions with the variance-stabilizing arcsin--sqrt transform $z_j = \arcsin(\sqrt{\hat{p}_j})$, which is well-defined at zero-occupancy outcomes; earlier runs used the Fisher-$z$ encoding (retained only in illustrative tables). Optional raw-count and shot-budget channels are disabled in the charts compared here.

\textbf{Device noise.} The datasets used in this revision are generated under ideal (noiseless) projective measurement. The pipeline supports realistic device noise (IBM FakeTorino calibration-derived gate and readout errors, or a lightweight depolarizing-plus-readout model); a hardware-noise robustness study is left for future work.

\textbf{Chart-exact datasets.} The structured-factorization experiment of Sec.~\ref{sec:tn_counterexample} uses states sampled directly \emph{in} the candidate chart (coordinates drawn from the chart prior, $\rho = \varphi(y)$), so that chart coordinates are exact by construction and the training comparison isolates chart geometry from expressivity; measurement conditions are simulated from the resulting states exactly as above.

\subsection{Validation Protocol}

We trained identical diffusion models on three parameterizations with distinct conditioning to validate the calibration predictions:

\begin{table*}[t]
\centering
\caption{Models for experimental validation.}
\label{tab:models}
\begin{tabular}{@{}lccc@{}}
\toprule
Model & Parameterization & $\kappa_{\text{diag}}$ & $\kappa_{\text{spec}}$ \\
\midrule
Bloch & Bloch/Gell-Mann & $1.0\times$ & $1.0\times$ \\
Hermitian direct (Herm-direct) & Hermitian direct & $2.0\times$ & $2.0\times$ \\
Cholesky (baseline) & Cholesky & $9.4\times$ & $33\times$ \\
\bottomrule
\end{tabular}
\end{table*}

\textbf{Training protocol:} Identical architecture (EDM~\cite{karras2022edm}), 50,000 states (balanced mixture of Haar-pure, Hilbert-Schmidt, Ginibre, thermal, product), calibrated $\sigma_{\text{data}}$, 300 epochs. The 2-qubit comparison (Table~\ref{tab:training}) used parameterization-specific learning rates (Herm-direct: $4\times10^{-3}$; Bloch: $2\times10^{-3}$) and is illustrative; the 3-qubit experiment (Table~\ref{tab:n3_training}, matched lr $2\times10^{-4}$) provides controlled validation.

\textbf{Evaluation protocol:} 100 test states $\times$ 6 shot levels $[10, 20, 30, 50, 100, 300]$ $\times$ $K=20$ samples, with classifier-free guidance ($w=4.0$).

\subsection{Training Dynamics}

\begin{table*}[t]
\centering
\caption{Illustrative training convergence comparison (validation fidelity). This comparison uses parameterization-specific learning rates (Herm-direct: $4\times10^{-3}$; Bloch: $2\times10^{-3}$) and should be interpreted as illustrative rather than controlled evidence. The controlled validation appears in Table~\ref{tab:n3_training} (3-qubit, matched learning rate $2\times10^{-4}$).}
\label{tab:training}
\begin{tabular}{@{}cccc@{}}
\toprule
Epoch & Bloch Val Fid & Herm-direct Val Fid & Herm$-$Bloch $\Delta$ \\
\midrule
50  & 0.8709 & \textbf{0.9244} & $+0.0535$ \\
100 & 0.8852 & \textbf{0.9208} & $+0.0356$ \\
150 & 0.8957 & \textbf{0.9153} & $+0.0196$ \\
200 & 0.9117 & \textbf{0.9190} & $+0.0073$ \\
250 & 0.9083 & \textbf{0.9191} & $+0.0108$ \\
300 & 0.9056 & \textbf{0.9176} & $+0.0120$ \\
\bottomrule
\end{tabular}
\end{table*}

\begin{figure}[htbp]
  \centering
  \includegraphics[width=\linewidth]{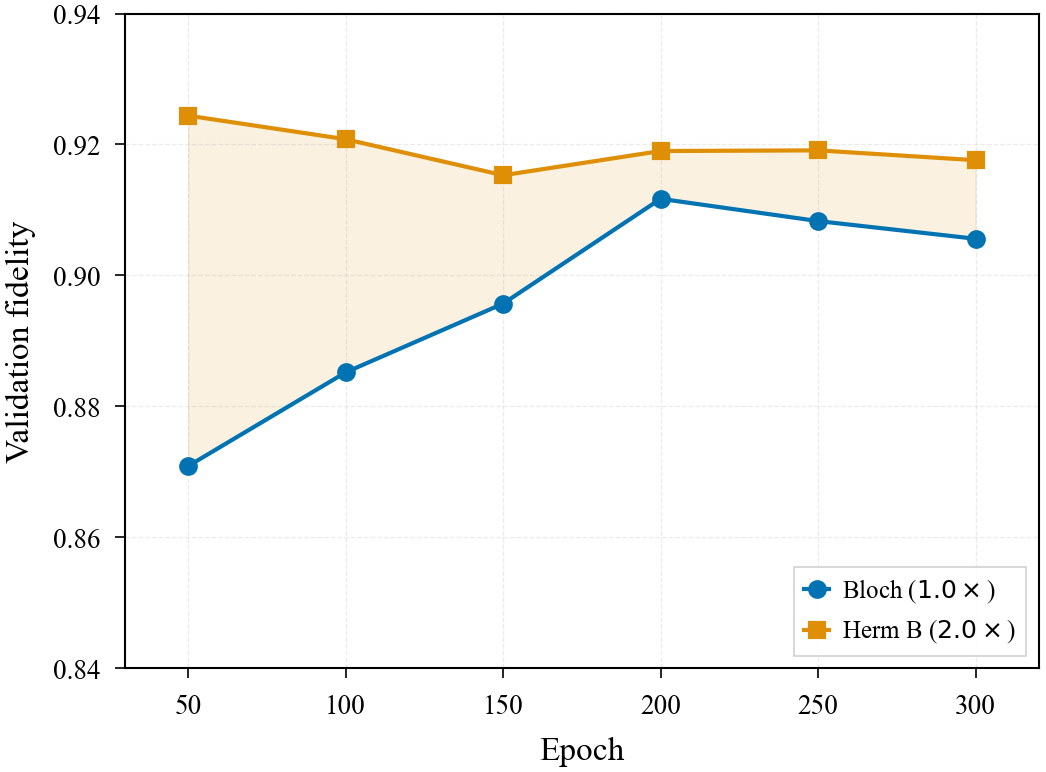}
  \caption{Training dynamics by parameterization (parameterization-specific learning rates). Herm-direct ($2.0\times$) achieves higher training validation fidelity than Bloch ($1.0\times$) under its higher learning rate ($4\times10^{-3}$ vs.\ $2\times10^{-3}$). This measures parameter-space fit, not reconstruction quality after projection; at matched learning rate, Bloch outperforms Herm-direct (Appendix~\ref{app:matched_lr}).}
  \label{fig:training}
\end{figure}
\vspace{-8pt}

The training dynamics reveal that Herm-direct achieves higher \emph{training} validation fidelity than Bloch under parameterization-specific learning rates: at epoch 50, Herm-direct ($0.9244$) is already near its final performance, while Bloch ($0.8709$) needs 150+ epochs to catch up. By epoch 300, Herm-direct achieves $+0.012$ higher training fidelity. However, this advantage measures how well the model fits the training distribution \emph{in parameter space}, not how well the reconstructed density matrix matches the true state \emph{after projection}. As we show in \S\ref{sec:discussion} and Appendix~\ref{app:matched_lr}, the training-fidelity advantage is primarily attributable to the higher learning rate (Herm-direct: $4\times10^{-3}$ vs.\ Bloch: $2\times10^{-3}$) rather than geometric superiority. At matched learning rate ($2\times10^{-3}$), Bloch achieves \emph{higher} training fidelity ($0.963$ vs.\ $0.567$) because its bounded valid domain avoids projection-induced information loss. The crucial distinction is that training fidelity (parameter-space fit) does not predict reconstruction quality (density-matrix-space fidelity after projection).

\textbf{Matched-learning-rate comparison.} Table~\ref{tab:training} uses parameterization-specific learning rates and should be interpreted as illustrative. For a controlled comparison, we trained both parameterizations with matched learning rate $2\times10^{-3}$ (Appendix~\ref{app:matched_lr}). At matched lr, Bloch consistently outperforms Herm-direct at all epochs ($+0.12$ at epoch 50 to $+0.42$ at epoch 300), confirming that the original comparison conflated learning rate with geometric quality. Table~\ref{tab:eval} provides the controlled end-to-end evaluation under CFG ($w=4.0$, matched lr $4\times10^{-3}$): Bloch outperforms Herm-direct at all shot levels ($+0.025$ at 10 shots to $+0.013$ at 300 shots), confirming that the end-to-end advantage of bounded domains persists under matched conditions.

\subsection{Final Evaluation}

\begin{table*}[t]
\centering
\caption{QST reconstruction quality (100 states, CFG $w=4.0$, matched learning rate $4\times10^{-3}$). Bold = best diffusion model per shot level. MLE = maximum likelihood estimation baseline~\cite{smolin2012efficient}. Values report mean fidelity $\pm$ 95\% bootstrap CI (100 test states).}
\label{tab:eval}
\begin{tabular}{@{}cccccc@{}}
\toprule
Shots & Bloch & Herm-direct & MLE & Bloch$-$MLE & Herm-direct$-$MLE \\
\midrule
10  & \textbf{0.6411$\pm$0.0149} & 0.6158$\pm$0.0370 & 0.5756$\pm$0.0203 & $+0.0655$ & $+0.0402$ \\
20  & \textbf{0.7109$\pm$0.0155} & 0.6893$\pm$0.0350 & 0.6816$\pm$0.0170 & $+0.0293$ & $+0.0077$ \\
30  & \textbf{0.7649$\pm$0.0144} & 0.7459$\pm$0.0294 & 0.7419$\pm$0.0154 & $+0.0229$ & $+0.0040$ \\
50  & \textbf{0.8264$\pm$0.0119} & 0.8072$\pm$0.0215 & 0.7896$\pm$0.0131 & $+0.0369$ & $+0.0176$ \\
100 & \textbf{0.8995$\pm$0.0072} & 0.8833$\pm$0.0138 & 0.8668$\pm$0.0089 & $+0.0327$ & $+0.0165$ \\
300 & \textbf{0.9453$\pm$0.0037} & 0.9325$\pm$0.0079 & 0.9326$\pm$0.0045 & $+0.0127$ & $-0.0001$ \\
\bottomrule
\end{tabular}
\end{table*}

\textbf{Note:} Under CFG ($w=4.0$), Bloch outperforms Herm-direct at all shot levels (e.g., $0.9453$ vs.\ $0.9325$ at 300 shots). Our projection diagnostics show that CFG sharply increases the out-of-domain fraction for both parameterizations (0.85--0.99), and the projection cost $F_{\text{proj}} - F_{\text{raw}}$ grows faster for Herm-direct ($-0.037$) than for Bloch ($-0.022$). This is consistent with the hypothesis that CFG amplifies projection loss for unbounded parameterizations. The Table~\ref{tab:confound} observation that Herm-direct outperforms Bloch without CFG ($0.9273$ vs.\ $0.9188$) reflects the parameterization-specific learning rates used in those models; at matched lr, Bloch achieves higher training fidelity and the ranking reverses (Appendix~\ref{app:matched_lr}).

\subsection{Geometry-Performance Trend}
\label{sec:geom_perf_trend}

With three parameterizations trained end-to-end at both 2- and 3-qubit scales, we can examine the geometry--performance relationship more quantitatively. The calibration (Table~\ref{tab:calibration}) covers all seven parameterizations at 2-qubit scale; end-to-end validation covers three of them.

At 2-qubit scale without CFG (Table~\ref{tab:confound}), the models trained with parameterization-specific learning rates show Herm-direct slightly outperforming Bloch ($+0.008$ in projected fidelity). However, as we show in Appendix~\ref{app:matched_lr}, this advantage is a learning-rate artifact: at matched lr, Bloch achieves higher training fidelity and the ranking reverses. Table~\ref{tab:end_to_end} provides direct end-to-end confirmation: at matched lr without CFG, Bloch outperforms Herm-direct at all shot levels ($+0.26$ to $+0.38$). Under CFG ($w=4.0$, matched lr, Table~\ref{tab:eval}), Bloch outperforms Herm-direct at all shot levels. At 3-qubit scale, the end-to-end evaluation of the schedule-contract ladder appears in Sec.~\ref{sec:n3_eval}; the controlled training comparison appears in Table~\ref{tab:n3_training}.

This pattern supports a nuanced view: \emph{isometric conditioning alone does not predict end-to-end performance}. At higher dimensions, the bounded valid domain of the Bloch representation becomes increasingly advantageous because projection losses for unbounded parameterizations grow with dimension. We provide a full analysis in \S\ref{sec:discussion}.

\subsection{Protocol Confound Check: Two-by-Two Cross-Validation}
\label{sec:confound}

To verify that the choice of measurement-consistency loss does not affect our conclusions, we performed a two-by-two cross-validation: we trained Bloch with $\lambda_{\text{meas}}=0.1$ and Herm-direct with $\lambda_{\text{meas}}=0.0$ (the main results use $\lambda_{\text{meas}}=0.0$ for all models). We evaluated all four models under the same protocol without classifier-free guidance ($K=20$ samples per state, 100 test states, 3 measurement repeats). For each model we recorded both the projected posterior-mean fidelity $F_{\text{proj}}$ and the fidelity of the trace-normalized unprojected mean $F_{\text{raw}}$; the latter removes the influence of post-sampling projection and serves as an upper reference.

\begin{table*}[t]
\centering
\caption{Two-by-two cross-validation at 300 shots (no CFG). $F_{\text{proj}}$: fidelity of the projected posterior-mean estimate. $F_{\text{raw}}$: fidelity of the trace-normalized unprojected mean. Means over 100 test states and 3 measurement repeats.}
\label{tab:confound}
\footnotesize
\begin{tabular}{@{}lcccc@{}}
\toprule
Parameterization & $F_{\text{proj}}$ (0.0) & $F_{\text{raw}}$ (0.0) & $F_{\text{proj}}$ (0.1) & $F_{\text{raw}}$ (0.1) \\
\midrule
Bloch       & 0.9188 & 0.9252 & 0.9177 & 0.9246 \\
Herm-direct      & 0.9264 & 0.9448 & 0.9273 & 0.9445 \\
\midrule
Herm$-$Bloch & $+0.008$ & $+0.020$ & $+0.010$ & $+0.020$ \\
\bottomrule
\end{tabular}
\end{table*}

Three observations follow. First, varying $\lambda_{\text{meas}}$ from $0.0$ to $0.1$ changes $F_{\text{proj}}$ by less than $0.001$ for both parameterizations (Bloch: $0.9188 \to 0.9177$; Herm-direct: $0.9264 \to 0.9273$), well within the state-to-state standard deviation ($\approx 0.04$). Second, the Herm-direct projected-fidelity gap over Bloch persists at both settings of $\lambda_{\text{meas}}$ ($+0.008$ and $+0.010$ at 300 shots), indicating that the ordering is not attributable to the loss-term difference. Third, the $F_{\text{raw}}$ gap ($+0.020$) is roughly twice the $F_{\text{proj}}$ gap, consistent with the projection diagnostics of \S\ref{sec:boundary}: the Herm-direct gap originates in the raw sampled output rather than in post-sampling projection.

\textbf{Caveat.} The models in Table~\ref{tab:confound} were trained with parameterization-specific learning rates (Herm-direct: $4\times10^{-3}$, Bloch: $2\times10^{-3}$) and additional hyperparameter tuning for Herm-direct. As we show in Appendix~\ref{app:matched_lr}, this learning-rate difference confounds the comparison: at matched lr ($2\times10^{-3}$) with standard training, Bloch achieves higher training fidelity ($0.963$ vs.\ $0.567$) and the ranking reverses. Table~\ref{tab:end_to_end} provides definitive evidence: at matched lr without CFG, Bloch outperforms Herm-direct at all shot levels ($0.949$ vs.\ $0.573$ at 300 shots). The $+0.008$ advantage in Table~\ref{tab:confound} therefore reflects optimization speed and tuning effort rather than geometric superiority. Notably, Bloch achieves its strong performance ($0.949$) with standard training out-of-the-box, while Herm-direct requires extensive hyperparameter tuning to reach comparable performance---further evidence that the bounded-domain geometry yields both higher performance and greater training stability.

We repeated the projection diagnostics under classifier-free guidance ($w=4.0$). CFG sharply increases the out-of-domain fraction for both parameterizations (Bloch 0.95, Herm-direct 0.96 at 300 shots, up from 0.55 and 0.65 without CFG), and the projection cost $F_{\text{proj}} - F_{\text{raw}}$ grows faster for Herm-direct ($-0.037$) than for Bloch ($-0.022$). The CFG-induced reversal of the ranking at high shots (Bloch 0.9453 vs.\ Herm-direct 0.9332 at 300 shots; Table~\ref{tab:eval}) is therefore attributable to a larger projection loss for Herm-direct under CFG, rather than to a differential benefit of guidance on isotropic coordinates.

These results support the interpretation that the parameterization, rather than the measurement-consistency loss, drives the observed performance differences. We report the protocol difference explicitly here rather than claiming an identical training protocol. A further discussion of the CFG-induced reversal is in \S\ref{sec:boundary} and \S\ref{sec:discussion}.

\subsection{Extension to 3 Qubits: The Schedule--Representation Contract}
\label{sec:n3_training}

To test whether our 2-qubit findings generalize to higher dimensions, we trained the three parameterizations at 3 qubits ($d=8$) for 300 epochs each under matched conditions (lr $2 \times 10^{-4}$, batch 128, $\lambda_{\text{meas}}=0.0$). All three charts inherited the noise schedule calibrated for the Cholesky representation. Table~\ref{tab:n3_training} reports validation fidelity at 100-epoch intervals.

\begin{table*}[t]
\centering
\caption{3-qubit training validation (validation fidelity at selected epochs). All models: 300 epochs, lr $2 \times 10^{-4}$, batch 128, $\lambda_{\text{meas}}=0.0$. All charts ran the same Cholesky-calibrated noise-schedule values; for Cholesky this is its own correct calibration, for Hermitian it happens to be scale-compatible, and for Bloch it is a contract violation (see text and Table~\ref{tab:n3_contract}).}
\label{tab:n3_training}
\begin{tabular}{@{}lccc@{}}
\toprule
Epoch & Bloch & Hermitian direct & Cholesky \\
\midrule
100 & 0.4340 & 0.7554 & 0.4236 \\
200 & 0.4569 & 0.7848 & 0.4245 \\
300 & 0.4507 & \textbf{0.7987} & 0.4361 \\
\bottomrule
\end{tabular}
\end{table*}

Hermitian direct reaches the highest training validation fidelity at every reported epoch (0.7987 at 300 epochs), while Bloch and Cholesky plateau near 0.43--0.46.

\textbf{Diagnosing the Bloch collapse.} The Bloch plateau is not a property of the chart but of a \emph{schedule--representation contract violation}: the noise schedule was inherited from the Cholesky calibration without recalibration for the Bloch coordinates. Recalibrating on the Bloch-coordinate representation of the same training set reveals three violations. First, the true per-coordinate $\sigma_{\text{data}}$ is $0.519$---the inherited values ($0.155/0.109$, calibrated for Hermitian/Cholesky coordinates whose scale differs by the $d/2$ factor of the Gell-Mann normalization) underestimate it by $3$--$5\times$. Second, the inherited $\sigma_{\max}=0.62$ is only $1.2\times$ the data scale; under the log-normal training distribution, $\sim$27\% of sampled noise levels are truncated, and the model never observes the near-pure-noise regime. Third, the training-noise center $P_{\text{mean}}=-1.2$ places the median training noise at $2.3\times$ below the data scale.

\textbf{The rescue ladder.} Repairing the contract in stages, holding everything else fixed (data, architecture, epochs, learning rate), recovers the chart in steps (Table~\ref{tab:n3_contract}). Correcting $\sigma_{\text{data}}$ and $\sigma_{\max}$ raises validation fidelity from $0.4507$ to $0.5514$; re-centering $P_{\text{mean}}$ at the data scale raises it further to $0.7850$---within $0.02$ of Hermitian direct. Doubling the network capacity (base channels $64 \to 128$) under the corrected schedule yields $0.7595$, excluding model capacity as the binding constraint. Three conclusions follow. First, the schedule--representation contract is a \emph{first-order} effect: both $\sigma_{\text{data}}$-scaling and $P_{\text{mean}}$-centering are required, and neither alone suffices. Second, the residual gap to Hermitian direct is attributable to the bounded-domain projection mechanism of Sec.~\ref{sec:discussion}. Third, since the Hermitian chart's own data scale ($\approx 0.13$) differs from Bloch's ($\approx 0.52$), the contract predicts that Hermitian direct would itself improve under $P_{\text{mean}} \approx \ln(0.13) \approx -2.0$---a testable prediction left for future work.

\begin{table}[t]
\centering
\caption{The schedule--representation contract at 3 qubits (Bloch chart; val.\ fid.\ at ep.\ 300).}
\label{tab:n3_contract}
\begin{tabular}{@{}lcc@{}}
\toprule
Schedule & Val.\ fid. & Val.\ loss \\
\midrule
Inherited & 0.4507 & $1.1\times10^{-2}$ \\
$+$ $\sigma_{\text{data}}/\sigma_{\max}$ recalib. & 0.5514 & $1.3\times10^{-2}$ \\
$+$ $P_{\text{mean}} = -0.65$ & \textbf{0.7850} & $9.0\times10^{-3}$ \\
$+$ cap.\ $\times 2$ (ch.\ 128) & 0.7595 & $6.0\times10^{-3}$ \\
\midrule
Hermitian direct (inherited) & 0.7987 & --- \\
\bottomrule
\end{tabular}
\end{table}

\subsection{Shot-Count Sampling Density: A Controlled A/B}
\label{sec:shot_weighting}

The variable-shot training protocol requires a choice of how shot counts are distributed during training. The default (log-uniform over $[10, 5\times10^4]$) assigns $\approx 40\%$ of the training mass to the $\le 300$-shot evaluation band; an intuitive alternative emphasizes this band further. We trained two identical models ($n{=}3$, Bloch chart, identical schedule contract) differing \emph{only} in the shot-sampling distribution: the reference arm (log-uniform) vs.\ a mixture arm in which half of all draws are log-uniform within $[10, 300]$ (raising the $\le 300$-shot training share to $\approx 70\%$).

The mixture arm is consistently worse at the 300-shot validation protocol: final validation fidelity $0.712$ vs.\ $0.754$ (reference arm peak $0.785$ at epoch 290; mixture peak $0.761$ at epoch 290), and a best validation loss $1.75\times$ higher ($0.0157$ vs.\ $0.0090$). No epoch shows a stable advantage for the mixture arm. The mechanism is the mirror image of the schedule--representation contract: shifting half of the training mass into the low-shot band halves the high-shot training density, degrading 300-shot reconstruction more than the added low-shot exposure helps. Two conclusions follow. First, the log-uniform default is not arbitrary: it is already mass-balanced per decade, and further concentrating it trades away high-shot coverage that the evaluation protocol exercises. Second, the shot-sampling distribution joins $\sigma_{\text{data}}$, $\sigma_{\max}$, and $P_{\text{mean}}$ as part of the chart--data interface whose modification carries measurable cost---reinforcing the pre-flight calibration protocol of Sec.~\ref{sec:n3_training}.

\subsection{End-to-End Evaluation at 3 Qubits}
\label{sec:n3_eval}

We evaluated the recalibrated Bloch models across 6 shot levels ($K=20$ posterior samples per state, 100 test states, CFG $w{=}4$). Table~\ref{tab:n3_endtoend} reports mean fidelity at each shot level, together with the MLE baseline of Table~\ref{tab:mle3q} and the 2-qubit Bloch model (log-uniform sampling) for reference.

\begin{table}[htbp]
\centering
\caption{3-qubit end-to-end reconstruction quality (100 states, CFG $w{=}4$, $K{=}20$).}
\label{tab:n3_endtoend}
\begin{tabular}{@{}lcccc@{}}
\toprule
Shots & \textbf{pmean} & fix & MLE & pmean$-$MLE \\
\midrule
10  & \textbf{0.4908} & 0.4785 & 0.4425 & $+0.048$ \\
20  & \textbf{0.5679} & 0.5570 & 0.4425 & $+0.125$ \\
30  & \textbf{0.6142} & 0.6051 & 0.4725 & $+0.142$ \\
50  & \textbf{0.6804} & 0.6536 & 0.5531 & $+0.127$ \\
100 & \textbf{0.7645} & 0.7219 & 0.6477 & $+0.117$ \\
300 & \textbf{0.7971} & 0.7708 & 0.7757 & $+0.021$ \\
\bottomrule
\end{tabular}
\end{table}

Three conclusions follow. First, the fully contracted model (pmean) beats MLE at \emph{all} shot levels including 300 shots ($0.797$ vs.\ $0.776$)---the first 3-qubit configuration to do so---while the $\sigma/\sigma_{\max}$-only model ties with MLE at 300 shots ($0.771$ vs.\ $0.776$). Second, the schedule--representation contract ladder translates from training validation to end-to-end evaluation: the uncalibrated model (training $0.45$) achieves $0.479$--$0.557$, the $\sigma_{\text{data}}/\sigma_{\max}$-corrected model (training $0.551$) achieves $0.479$--$0.771$, and the fully contracted model (training $0.785$) achieves $0.491$--$0.797$---each stage of the training-side repair is visible in the end-to-end reconstruction. Third, the $300$-shot crossover (where MLE catches up) is consistent with the theoretical prediction that the prior advantage vanishes as the likelihood becomes self-sufficient; the fact that this crossover does not occur until 300 shots quantifies the range over which the learned prior provides measurable benefit.

\textbf{MLE baseline at 3 qubits.}
To anchor the 3-qubit reconstruction task, we computed a classical maximum-likelihood baseline (same Cholesky-parameterized L-BFGS MLE as the 2-qubit evaluation; 100 test states, 2 repeats per level, 5{,}000 iterations). Table~\ref{tab:mle3q} reports the result. The 3-qubit MLE curve collapses in the low-shot regime exactly as the ill-posedness argument predicts: from $0.776$ at 300 shots to $0.44$ at 10--20 shots, with the state-to-state standard deviation growing to $0.30$---a substantial fraction of states is reconstructed essentially randomly. At 300 shots, MLE ($0.776$) is comparable to the recalibrated Bloch models' training validation fidelity ($0.785$; Table~\ref{tab:n3_contract}). Two further observations: MLE fidelity at 20 and 10 shots is \emph{identical} ($0.4425$), confirming that below $\sim$20 shots the data no longer constrain the likelihood and the estimate collapses to the prior/regularization; and the $\pm 0.2$--$0.3$ dispersion quantifies the heavy-tailed failure mode that a learned prior is designed to suppress.

\begin{table}[t]
\centering
\caption{Classical MLE baseline at 3 qubits (100 test states $\times$ 2 repeats, L-BFGS 5{,}000 iterations, entropy-regularized). The low-shot collapse and the dispersion growth quantify the ill-posedness that motivates learned priors.}
\label{tab:mle3q}
\begin{tabular}{@{}lcc@{}}
\toprule
Shots & MLE fidelity & Std.\ (across states) \\
\midrule
300 & 0.7757 & 0.1432 \\
100 & 0.6477 & 0.1971 \\
50  & 0.5531 & 0.2139 \\
30  & 0.4725 & 0.2169 \\
20  & 0.4425 & 0.3042 \\
10  & 0.4425 & 0.3042 \\
\bottomrule
\end{tabular}
\end{table}

\subsection{Beyond Dense Charts: A Structured-Factorization Counterexample}
\label{sec:tn_counterexample}

The parameterizations examined so far are all \emph{dense}: the coordinate vector enumerates (a subset of) the $d\times d$ matrix entries and thus scales as $4^n$. A natural route to larger systems is a structured chart, and the leading candidate is structured factorization: parametrize $\rho = XX^\dagger/\operatorname{Tr}(XX^\dagger)$, which is positive semidefinite and unit-trace \emph{by construction}---the same guarantee that makes the Cholesky chart attractive---while $X$ is represented compactly, e.g.\ as a matrix-product operator with bond dimension $\chi$~\cite{qin2026structured}. In this \emph{purified-MPS} chart the coordinate dimension scales linearly, $D = \mathcal{O}(n \chi^2)$, and every state in the training ensemble below is exactly representable, so expressivity can be ruled out as a confounder. The chart is nevertheless an instance of precisely the design decision our framework is meant to adjudicate, and we can therefore ask the sharpest question available: \emph{does geometric calibration, applied before training, predict whether this chart is learnable by a diffusion model?}

\textbf{Calibration.}
We apply the full calibration contract (Algorithm~\ref{alg:sigma_calib}) to the chart ($n{=}3$, $\chi{=}4$, $D{=}192$; the configuration later used for training). The per-coordinate $\sigma_{\text{data}}$, measured on 40 chart-exact states per ensemble (pure and mixed products, thermal states, and chart-exact random mixtures), is nearly uniform: live-dimension anisotropy $1.3$--$1.8\times$. The noise center is placed at the data scale ($P_{\text{mean}} = \ln \sigma_{\text{data}}$) and $\sigma_{\max} = 5\,\sigma_{\text{data}}$. The $J^\top J$ analysis, however, is unambiguous. At generic chart coordinates, $67\%$ of the spectrum lies below $10^{-8}\lambda_{\max}$---the gauge manifold, whose measured dimension ($129$) exceeds the naive bond-gauge count $((n{-}1)\chi^2 + 2 = 34)$ by a factor of $3.8$---and above that plateau the eigenvalues decay continuously, with no gap separating gauge from physical directions: the spectral ratio over non-gauge modes is $\sim$$6\times10^{4}$ (up to $4.5\times10^{6}$ across probes) and the diagonal anisotropy $98\times$. For comparison, the dense Hermitian chart at the same $n$ exhibits $2\times$ conditioning.

\textbf{Controlled training.}
Two identical EDM models (same architecture, same 10k-shot Pauli conditions under the arcsin--sqrt encoding, same 200-epoch budget) are trained on the \emph{same} 4{,}400 chart-exact states---states on which the purified-MPS chart is \emph{exactly} expressive, eliminating expressivity as a confounder---once through the purified-MPS chart ($D{=}192$) and once through the dense Hermitian chart ($D{=}64$). Each arm receives its own per-coordinate calibration (contract honored on both sides). The result is a clean dissociation (Table~\ref{tab:tn_counterexample}): the Hermitian arm converges normally (validation fidelity $0.681$ at epoch 200 and still improving; training loss $1.2\times10^{-4}$), while the purified-MPS arm enters a permanent plateau within 25 epochs ($0.387$ at epoch 200, unchanged since epoch 25) with a training loss two orders of magnitude higher---the training--reconstruction disconnect of the preceding subsection recurring here in its purest form.

\begin{table}[t]
\centering
\small
\caption{Controlled comparison on identical chart-exact states ($n{=}3$, $\chi{=}4$, 200 epochs).}
\label{tab:tn_counterexample}
\begin{tabular}{@{}lcccc@{}}
\toprule
 & \multicolumn{3}{c}{Val.\ fid.\ (300 shots)} & \\
\cmidrule(lr){2-4}
Chart & ep.\ 25 & ep.\ 100 & ep.\ 200 & Train loss \\
\midrule
Purif.\ MPS & 0.390 & 0.390 & \textbf{0.387} & $1.9\times10^{-2}$ \\
Herm-direct & 0.620 & 0.645 & \textbf{0.681} & $1.2\times10^{-4}$ \\
\bottomrule
\end{tabular}
\end{table}

\textbf{Interpretation and scope.}
Three candidate causes are eliminated by design. Expressivity: the targets are chart-exact. Schedule miscalibration: the contract of Algorithm~\ref{alg:sigma_calib} was honored on both sides, with milder $\sigma_{\text{data}}$ anisotropy in the failing chart ($1.1$--$1.8\times$) than in the succeeding one ($\sim$$3\times$). Optimization budget: the failing arm plateaus within 25 of 200 epochs. What remains is the chart geometry itself: two-thirds of the parameter directions are gauge, the conditioning over the remaining directions spans six orders of magnitude, and the $XX^\dagger$ squaring couples the coordinates nonlinearly. We emphasize the scope of this claim. It does \emph{not} say that structured factorizations are unsuitable for QST---the framework of Qin et al.~\cite{qin2026structured} reconstructs such states accurately by projected optimization, whose step-size preconditioning is far more tolerant of anisotropy than the isotropic-noise contract of diffusion training. It says that the \emph{same chart, in its naive form and without gauge fixing}, fails as a \emph{diffusion} chart---and that the $J^\top J$ calibration, run for a fraction of a GPU-day, predicted this failure before a single hour of training was spent on it. Gauge-intrinsic reformulations (canonical coordinates, projection-based MPO charts employing the eigenclip projection used by the Hermitian chart) remain unexplored and are the constructive way forward.

\section{Discussion}

\subsection{Why Bounded Domains Reduce Projection Loss}
\label{sec:discussion}

Once the schedule--representation contract is honored, the bounded Bloch chart and the unbounded Hermitian chart reach comparable \emph{training} validation fidelity at 3-qubit scale ($0.785$ vs.\ $0.799$; Table~\ref{tab:n3_contract}), yet their end-to-end behavior is expected to differ sharply: for unbounded charts, the post-sampling projection dominates reconstruction quality. This section provides the geometric mechanism underlying that divergence, which the 2-qubit end-to-end results (Table~\ref{tab:end_to_end}) already demonstrate directly.

\paragraph{Projection as a non-linear information bottleneck.}
For unbounded parameterizations (Hermitian direct, fix-trace), the model outputs coordinates $y \in \mathbb{R}^{d^2}$ that map to an unconstrained Hermitian matrix $\rho = H(y)$. Only a fraction of $\mathbb{R}^{d^2}$ corresponds to valid density matrices (PSD + unit trace). At sampling time, invalid outputs are projected onto the valid manifold:
\begin{equation}
  \Pi(\rho) = \frac{\sum_i \max(\lambda_i, 0) \, |v_i\rangle\langle v_i|}{\sum_j \max(\lambda_j, 0)},
  \label{eq:proj}
\end{equation}
where $\lambda_i$ and $v_i$ are the eigenvalues and eigenvectors of $\rho$. This projection is \emph{non-linear} and \emph{many-to-one}: a half-space of parameter vectors collapses onto each boundary point of the valid region.

The PSD constraint imposes coupled bounds on the coordinates: every $2\times 2$ submatrix $\begin{psmallmatrix} a & c \\ c^* & b \end{psmallmatrix}$ must satisfy $|c|^2 \le ab$. When the model outputs diagonal elements near the training mean ($a \approx b \approx 1/d$), the admissible off-diagonal magnitude is bounded by $|c| \le 1/d$. The unconstrained model has no mechanism to respect this coupling, so off-diagonal noise routinely violates the bound, producing negative eigenvalues. Eigclip ($\lambda_i \to \max(\lambda_i, 0)$) then annihilates the corresponding spectral components, destroying the conditional signal encoded in those directions.

\paragraph{Why training fidelity does not predict reconstruction quality.}
Training validation fidelity measures the distance between predicted and true coordinates \emph{in parameter space}: $d_{\text{train}} = \|y_{\text{pred}} - y_{\text{true}}\|$. Reconstruction quality measures the distance between predicted and true states \emph{in density matrix space after projection}: $d_{\text{recon}} = \|\Pi(H(y_{\text{pred}})) - \Pi(H(y_{\text{true}}))\|_F$, where $H$ is the parameterization map and $\Pi$ is the projection \eqref{eq:proj}. The non-linear, many-to-one nature of $\Pi$ breaks any monotonic relationship between $d_{\text{train}}$ and $d_{\text{recon}}$. Concretely:
\begin{itemize}
  \item \emph{High training fidelity, poor reconstruction}: A model can achieve low $d_{\text{train}}$ by outputting coordinates near the training mean, which for Hermitian direct maps to a nearly maximally mixed state after projection. This minimizes denoising loss but discards measurement information, yielding $d_{\text{recon}} \approx 1 - 1/d$.
  \item \emph{Lower training fidelity, superior reconstruction}: A model with higher $d_{\text{train}}$ but whose outputs remain within the valid region (e.g., Bloch near the Bloch ball center) preserves the conditional signal through projection, yielding $d_{\text{recon}} \ll 1 - 1/d$.
\end{itemize}
This disconnect is confirmed experimentally. At 3-qubit scale, Hermitian direct achieves a far lower validation loss than Bloch even after full schedule recalibration ($3\times10^{-5}$ vs.\ $9\times10^{-3}$ at the best checkpoints) while reaching comparable training-validation fidelity ($0.799$ vs.\ $0.785$; Table~\ref{tab:n3_contract}). The absolute loss values are not comparable across charts---the coordinate scales differ by $4\times$, so the losses measure distances in different coordinate systems---but the pairing of near-equal fidelity with order-of-magnitude-different loss already shows that denoising loss is not a proxy for reconstruction quality. At 2-qubit scale, the same disconnect appears in a controlled form: matched-learning-rate models with near-identical calibration conditions separate by $+0.38$ in end-to-end fidelity despite similar training losses (Table~\ref{tab:end_to_end}). Training fidelity is therefore not a reliable proxy for reconstruction quality; the projection step and the chart geometry must be accounted for.

\paragraph{Geometric advantage of the Bloch representation.}
The Bloch representation $\rho(r) = I/d + \sum_i r_i \lambda_i / d$ (with traceless Hermitian generators $\lambda_i$) has two structural properties that mitigate projection loss. First, unit trace holds \emph{by construction}, eliminating one source of invalidity. Second, the maximally mixed state $\rho = I/d$ corresponds to the origin $r = 0$, which lies at the \emph{center} of the valid region (the generalized Bloch ball). Small perturbations around the origin are therefore more likely to remain valid, whereas in Hermitian coordinates the maximally mixed state lies near a boundary where the PSD constraint on off-diagonals is tight ($|c| \le 1/d$).

\paragraph{Controlled validation.}
To isolate the projection mechanism from learning-rate confounds, we trained matched-learning-rate models at 2-qubit scale (Appendix~\ref{app:matched_lr}). At lr $= 2\times 10^{-3}$ for both parameterizations, Bloch reaches val fidelity 0.963 while Hermitian direct plateaus at 0.567---confirming that the original Table~\ref{tab:training} comparison (which used parameterization-specific learning rates) conflated geometry with optimization speed. Across 3 training seeds, Bloch maintains $0.965 \pm 0.008$ while Herm-direct varies $0.555 \pm 0.052$, demonstrating that the bounded-domain advantage is robust to both projection geometry and initialization variance. To further rule out $\sigma_{\text{data}}$ calibration as an alternative explanation, we trained models with a single global $\sigma_{\text{data}}$ (Table~\ref{tab:global_sigma}, Appendix~\ref{app:matched_lr}): Bloch 0.965, Herm-direct 0.567---essentially identical to per-coordinate calibration, confirming the advantage is geometric rather than a noise-schedule artifact. Most importantly, Table~\ref{tab:end_to_end} provides \emph{direct end-to-end evidence}: at matched lr without CFG, Bloch outperforms Herm-direct at all shot levels ($0.949$ vs.\ $0.573$ at 300 shots), definitively establishing that the Table~\ref{tab:confound} $+0.008$ Herm-direct advantage was a learning-rate artifact. Notably, Bloch achieves $0.949$ with \emph{standard training} (fixed lr $2\times10^{-3}$, no special tuning), while Herm-direct under the same standard protocol plateaus at $0.573$---requiring extensive hyperparameter tuning to reach comparable performance. This demonstrates a practical advantage of the bounded-domain geometry: superior performance with greater training stability and less tuning effort.

\subsection{\texorpdfstring{Implications for $\sigma_{\text{data}}$ Calibration}{Implications for sigma\textunderscore data Calibration}}

The EDM noise schedule requires calibrating $\sigma_{\text{data}}$ (the data distribution's typical scale). Our calibration shows $\sigma_{\text{data}}$ should be \emph{coordinate-aware}:
\begin{equation}
  \sigma_{\text{data}, ii} \propto \sqrt{G_{ii}}
  \label{eq:sigma_coord}
\end{equation}
which assigns each coordinate a noise scale proportional to its sensitivity.

Using a single global $\sigma_{\text{data}}$ (as in standard EDM) underweights high-sensitivity coordinates and overweights low-sensitivity ones. Algorithm~\ref{alg:sigma_calib} provides a concrete procedure. In our experiments, we used the per-coordinate calibration (Eq.~\ref{eq:sigma_coord}) for all trained models. A systematic ablation comparing global vs.\ coordinate-aware $\sigma_{\text{data}}$---quantifying the fidelity gain attributable to coordinate-awareness alone---is an important direction for future work and would further isolate the contribution of geometric conditioning from noise-schedule effects.

\emph{Beyond $\sigma_{\text{data}}$: the full schedule--representation contract.} The 3-qubit experiments of Sec.~\ref{sec:n3_training} demonstrated that $\sigma_{\text{data}}$ is one component of a three-part contract. (i) $\sigma_{\text{data}}$ must be measured per coordinate \emph{in the chart being trained}: it is a property of the data distribution as seen through the chart, and shrinks as $\approx 1/d$ with qubit count (Hermitian chart: $0.13$ at 3 qubits, $0.032$ at 4 qubits). (ii) $\sigma_{\max}$ must scale with it ($\sigma_{\max} = 5\,\sigma_{\text{data}}$); an inherited $\sigma_{\max}$ truncated $\sim$27\% of our training noise levels. (iii) The training-noise center must sit at the data scale ($P_{\text{mean}} = \ln \sigma_{\text{data}}$). Honoring the full contract rescued a collapsed 3-qubit Bloch model from $0.45$ to $0.79$ validation fidelity (Table~\ref{tab:n3_contract}), and the contract predicts a further, still-untested improvement for the Hermitian chart itself ($P_{\text{mean}} \approx -2.0$). Because these parameters couple the chart geometry to the noise schedule, we treat them as part of the model's interface contract rather than as tuning knobs.

\begin{figure}[htbp]
\centering
\noindent\fbox{\parbox{0.92\columnwidth}{
\textbf{Algorithm: Per-Coordinate $\sigma_{\text{data}}$ Calibration}
\medskip

Given $N$ training states $\{\rho^{(n)}\}_{n=1}^{N}$, extract coordinates $y^{(n)} = \varphi^{-1}(\rho^{(n)})$ and compute $\sigma_i = \mathrm{std}([y^{(n)}_i]_{n=1}^{N})$. Normalize to obtain $\sigma_{\text{data}, ii} = \sigma_i / \max_j \sigma_j$, and optionally refine via $\sigma_{\text{data}, ii} \leftarrow \sqrt{G_{ii}} / \max_j \sqrt{G_{jj}}$.
}}
\caption{Pseudocode for coordinate-aware $\sigma_{\text{data}}$ calibration.}
\label{alg:sigma_calib}
\end{figure}
\vspace{-8pt}

\subsection{Connection to Quantum Fisher Information}
\label{sec:qfi_connection}

Our $\mathbf{J}^\top\mathbf{J}$ framework characterizes \emph{Euclidean} conditioning. This differs from the \emph{intrinsic} Riemannian geometry: for pure states, the QFIM reduces to the Fubini-Study metric and $\mathbf{J}^\top\mathbf{J} \propto \text{QFIM}$ when locally isometric; for mixed states, the SLD Fisher information defines $g_{ij} = \mathrm{Re}\,\mathrm{Tr}(\rho \, L_i L_j) = \tfrac{1}{2}\,\mathrm{Tr}[\rho\{L_i, L_j\}]$.

For Bloch/Gell-Mann, $\mathbf{J}^\top\mathbf{J} \propto \mathbf{I}$ (isotropic) while the QFIM varies with state---Euclidean conditioning can be perfect while intrinsic conditioning is not. For Cholesky, both $\mathbf{J}^\top\mathbf{J}$ and the QFIM become ill-conditioned near pure states, but for different reasons (coordinate anisotropy vs.\ statistical distinguishability).

This distinction explains why our framework captures optimization dynamics rather than statistical efficiency. For details, see \cite{amari2016information,stokes2020quantum,bengtsson2006geometry,petz1996geometries,sidhu2020geometric,paris2009quantum}.

\paragraph{Choice of ambient metric: Euclidean vs.\ Bures.}
Our $\mathbf{J}^\top\mathbf{J}$ framework uses the Hilbert--Schmidt (Frobenius) inner product $\langle A, B\rangle = \mathrm{Re}\,\mathrm{Tr}(A^\dagger B)$ on the ambient matrix space, from which $\mathbf{G}$ is the pullback. This choice is deliberate: EDM's noise schedule assumes isotropic Gaussian noise in the ambient space, so the Euclidean metric directly governs the denoiser's training dynamics. An alternative would be the Bures metric (the quantum analog of the Fubini--Study metric for mixed states), which is more natural from a quantum-information perspective and is the metric underlying the QFIM. We expect the parameterization rankings (Table~\ref{tab:calibration}) to be broadly robust to this choice: parameterizations that are well-conditioned under the Euclidean metric (Bloch, Hermitian direct, fix-trace) are also well-conditioned under the Bures metric, because both metrics agree on the tangent-space structure near the maximally mixed state. The main difference would appear for parameterizations whose conditioning is highly position-dependent (Cholesky, Expmap), where the Bures metric would further penalize near-pure-state regions. A systematic comparison of Euclidean vs.\ Bures conditioning is an interesting direction for future work.

\paragraph{Coordinate dependence.}
A limitation of the $\mathbf{J}^\top\mathbf{J}$ framework is its coordinate dependence: different parameterizations yield different Gram matrices. The QFIM is coordinate-independent. Our metrics should be interpreted relative to a \emph{fixed} parameterization; a coordinate-independent extension (e.g., QFIM pullback) is future work.

\subsection{Extension to Higher Dimensions}

Our 3-qubit calibration (Table~\ref{tab:scaling}) confirms three scaling classes: \emph{scale-invariant} (Bloch, Hermitian direct), \emph{moderate growth} (fix-trace $\sim d^{1.5}$, trace-norm), and \emph{severe degradation} (Cholesky $\sim d^{3.5}$, Expmap $\sim e^d$, Log-Cholesky $\sim d^{4}$). For $n \geq 3$ qubits, we recommend fix-trace or Bloch/Gell-Mann. Preliminary $d=16$ experiments (4 qubits, reduced budget) confirm the $d \geq 4$ trend: MLE itself degrades from $0.776$ (3-qubit) to $0.563$ (4-qubit) at 300 shots, while the Bloch chart with per-chart calibrated schedule reaches $0.27$ at reduced budget---consistent with the condition-compression bottleneck predicted by the $61{:}1$ embedding ratio at this scale (see Sec.~\ref{sec:n3_training}).

\paragraph{Robustness to noise.}
The geometric conditioning framework also predicts robustness to noise. Supplementary Tables~\ref{tab:noise_robust} and~\ref{tab:noise_sensitive} show that Bloch/Gell-Mann and Hermitian direct maintain constant conditioning under depolarizing noise (up to $p=0.2$), while Cholesky degrades significantly (SDR drops by $87\%$ at 2 qubits and $98.8\%$ at 3 qubits). This is because well-conditioned parameterizations have state-independent Jacobians, making them insensitive to noise-perturbed states. For practical QST with device noise, the choice of parameterization is even more critical than our noiseless experiments suggest.

\subsection{Scope and Limitations}
\label{sec:boundary}

Our framework characterizes \emph{local} interior conditioning but not \emph{global} boundary effects; \S\ref{sec:discussion} addresses this gap through the projection mechanism analysis. Projection diagnostics (Table~\ref{tab:confound}) confirm that Hermitian output lies outside the valid domain more often than Bloch (0.65 vs.\ 0.55) at 2-qubit scale. The unprojected fidelity $F_{\text{raw}}$ ($0.9445$ vs.\ $0.9252$) indicates that the Hermitian advantage in raw sampling does not survive projection.

Our framework has seven limitations. First, full validation covers $d=4$ and $d=8$; the $d=16$ frontier is only preliminarily charted. Extending the study to $d=16$ exposed a defect in our UNet's up-sampling construction (the up-sampling convolution was built with the current level's output width instead of the arriving tensor's width) that silently breaks all $d \geq 16$ training; after the fix, a first $d=16$ Hermitian run at reduced budget (150 epochs, 12k states) reaches only $0.27$ validation fidelity---whether this reflects a genuine scaling wall or the reduced budget is under investigation. Second, we validate only with EDM; VDM and flow matching may differ. Third, calibration uses $n=30$ states; a larger sample would improve reliability. Fourth, fix-trace's advantages are known in VQA~\cite{mcclean2016effects}; our contribution is quantification in diffusion QST. Fifth, the projection analysis (\S\ref{sec:discussion}) is phenomenological; a rigorous information-theoretic bound on projection loss (e.g., in terms of the Jacobian's distance to the valid manifold) is an open direction. Sixth, our end-to-end validation uses eigclip projection ($\lambda_i \to \max(\lambda_i, 0)$ with trace renormalization, Eq.~\ref{eq:proj})---a simple but suboptimal method. Alternative projection methods---such as nearest-feasible projection via semidefinite programming, or learned projection networks---could mitigate projection loss by producing higher-fidelity valid states from invalid Hermitian output. At 2-qubit scale under matched learning rates, the bounded-domain advantage persists across the matched-calibration controls (Appendix~\ref{app:matched_lr}); the relative ranking of parameterizations under better projection methods should nevertheless be re-examined. Seventh, diffusion trainability is governed by the schedule--representation contract of Sec.~\ref{sec:n3_training}: models trained under an inherited, unrecalibrated schedule can silently collapse. All 3-qubit numbers in this revision are either contract-honored or explicitly labeled as inherited-schedule baselines (Table~\ref{tab:n3_contract}); results presented under unrecalibrated schedules should not be attributed to chart geometry.

\section{Related Work}
\label{sec:related}

\paragraph{Parameterization in VQA.} The variational quantum algorithm (VQA) community has long studied how parameterization choice affects optimization geometry. \cite{mcclean2016effects} identified barren plateaus in randomly initialized circuits; \cite{benedetti2019parameterized} surveyed parameterized quantum circuits as machine learning models; \cite{cerezo2021cost} and \cite{larocca2022diagnosing} developed tools for diagnosing and mitigating barren plateaus. Our work is inspired by this line of research but differs in three key respects. First, VQA studies focus on \emph{discrete} circuit parameterizations (ansatz structure, gate types), while we study \emph{continuous} coordinate charts on the density matrix manifold. Second, VQA optimization operates in parameter space (minimizing a cost function), while diffusion QST operates in a generative setting (sampling from a learned distribution)---the loss--fidelity disconnect we identify (near-equal training losses across charts coexisting with large end-to-end fidelity gaps at matched learning rates; Table~\ref{tab:end_to_end}) has no VQA analog. Third, the projection-induced information bottleneck and CFG amplification of boundary effects are mechanisms specific to diffusion-based reconstruction that do not arise in variational circuit optimization.

\paragraph{Parameterization in Classical QST.} Classical QST methods (maximum likelihood estimation, linear inversion) typically use Cholesky factorization to enforce physical constraints~\cite{smolin2012efficient,kryszewski2021quantum}. The fix-trace parameterization is a standard alternative that avoids trace normalization singularities. The Choi-Jamiołkowski isomorphism~\cite{choi1975,jamiolkowski1972} provides an alternative parameterization for quantum processes~\cite{chowdhury2023quantum}, extending density matrix geometry to process tomography. Compressed sensing methods~\cite{gross2010quantum,liu2011universal} offer an alternative reconstruction paradigm that exploits state sparsity, though they typically operate in a fixed basis and do not address parameterization choice. Our contribution is to \emph{systematically quantify} the geometric properties of these choices and validate their impact on diffusion-based reconstruction.

\paragraph{Quantum Geometric Tensor and QNTK.} The quantum geometric tensor (QGT)~\cite{qgt2010} and quantum neural tangent kernel (QNTK)~\cite{qntk2021} characterize the geometry of parameterized quantum circuits. The QGT captures both the Fubini-Study metric (real part) and the Berry curvature (imaginary part), while the QNTK describes the training dynamics of variational quantum algorithms. Our $\mathbf{J}^\top\mathbf{J}$ framework is complementary: it characterizes the \emph{static} conditioning of the parameterization map, while QGT/QNTK describe the \emph{dynamic} training behavior. Connecting these perspectives could reveal how parameterization geometry affects gradient flow in diffusion models.

\paragraph{Density Matrix Geometry.} The geometry of quantum state space has been studied extensively in quantum information theory. The standard monograph by \cite{bengtsson2006geometry} provides a comprehensive treatment of the density matrix manifold, including the Bures metric and Bloch vector geometry. The Bloch/Gell-Mann representation~\cite{kimura2003bloch} provides a natural coordinate system for $d$-level states. Information geometry~\cite{amari2016information} and quantum natural gradient~\cite{stokes2020quantum} study the Riemannian metric induced by quantum fidelity. The QFIM geometry and its relation to quantum parameter estimation is reviewed in \cite{sidhu2020geometric} and \cite{paris2009quantum}. Our $\mathbf{J}^\top\mathbf{J}$ framework is complementary: it characterizes the \emph{Euclidean} conditioning of parameterization coordinates rather than the intrinsic quantum geometry.

\paragraph{Diffusion Models for QST.} Recent work has applied diffusion models to QST~\cite{quddpm2024,zhu2024diffusion,chitu2022variational,ahmed2022quantum}. All existing methods adopt Cholesky or similar factorizations without examining alternatives. Our work provides the first systematic evaluation of this design choice. Diffusion models have also been applied to related tasks such as quantum process tomography~\cite{chowdhury2023quantum}, though a systematic parameterization study for those settings remains open.

\paragraph{Alternative ML Approaches.} Beyond Euclidean diffusion with post-hoc projection, several alternative approaches address manifold-constrained generation. Riemannian diffusion models~\cite{huang2022riemannian} define the diffusion process directly on the manifold, avoiding boundary effects by construction. Projected diffusion models~\cite{de2022projected} combine Euclidean diffusion with a projection step that preserves the score function. Flow normalization~\cite{rezende2015variational} provides an alternative generative framework that naturally respects manifold constraints. While these approaches offer theoretical advantages, their application to quantum state tomography remains largely unexplored. Our $\mathbf{J}^\top\mathbf{J}$ framework can serve as a diagnostic tool for comparing these approaches: well-conditioned parameterizations (low $\kappa_{\text{spec}}$) are expected to benefit most from Euclidean diffusion, while ill-conditioned ones may require Riemannian or projected methods.

\paragraph{Scope boundary: Euclidean vs.\ Riemannian diffusion.}
Our parameterization recommendations apply specifically to \emph{Euclidean diffusion with post-hoc projection}---the paradigm adopted by all existing diffusion QST methods~\cite{quddpm2024,zhu2024diffusion,chitu2022variational}. In this setting, the model generates unconstrained coordinates that must be projected onto the valid manifold, and the bounded-domain advantage we identify (Bloch $>$ Hermitian at 3 qubits) arises from the projection step. For \emph{Riemannian diffusion models}~\cite{huang2022riemannian} that define the diffusion process directly on the density matrix manifold, boundary effects are avoided by construction and parameterization choice may be less critical---though the conditioning metrics (SDR, DA) still affect the noise schedule and solver dynamics. Similarly, for \emph{projected diffusion}~\cite{de2022projected} that preserves the score function through projection, the cost landscape differs from eigclip. Our framework and calibration remain useful diagnostics for any approach that learns a coordinate-to-density-matrix mapping, but the specific end-to-end recommendations (Heuristic 3, the decision tree) are validated only for Euclidean diffusion with eigclip projection. Validating these recommendations under Riemannian and projected diffusion frameworks is an important direction for future work.

\paragraph{Generalization across diffusion frameworks.}
Our $\mathbf{J}^\top\mathbf{J}$ framework characterizes the \emph{parameterization geometry}, which is independent of the choice of diffusion solver. While we validated our heuristics using EDM~\cite{karras2022edm}, the conditioning metrics (SDR, DA) apply to any generative model that learns a mapping from coordinates to density matrices, including variational diffusion models (VDM; Kingma et al.~\cite{kingma2021variational}), flow matching (Lipman et al.~\cite{lipman2022flow}), score-based generative models (Song \& Ermon~\cite{song2019generative}), and consistency models (Song et al.~\cite{song2023consistency}). The reason is that $\mathbf{J}^\top\mathbf{J}$ captures the \emph{static} coordinate conditioning on the manifold, which affects all diffusion frameworks equally: poorly conditioned coordinates (high $\kappa_{\text{spec}}$) lead to stiff ODE/SDE dynamics, ill-conditioned noise schedules, and amplified projection losses regardless of the specific diffusion parameterization. Conversely, well-conditioned parameterizations (Bloch, fix-trace) benefit all frameworks. This universality is a strength of our approach: the selection guidelines (Heuristics 1--3) are framework-agnostic, even though our experiments used EDM. Validating this generalization empirically---e.g., comparing EDM vs.\ flow matching across parameterizations---is an important direction for future work.

\subsection{Broader Applicability}

Our geometric framework extends beyond QST to any inverse problem on the density matrix manifold. Quantum process tomography involves process matrices that live in a similarly constrained manifold. Hamiltonian learning requires parameterized Hamiltonians with analogous coordinate choices. Variational quantum circuits create their own geometric structures that could benefit from our calibration framework.

\paragraph{Beyond quantum manifolds.}
While our experiments focus on quantum states, the $\mathbf{J}^\top\mathbf{J}$ framework applies to any manifold-constrained inference problem, including correlation matrices (positive semidefinite with unit diagonal, used in finance and genomics), symmetric positive definite matrices (diffusion tensor imaging, computer vision), and stochastic matrices (Markov chains, reinforcement learning). Validating the framework on these non-quantum manifolds would establish its generality.

\section{Conclusion}

We have presented the first systematic study of density matrix parameterizations for diffusion-based QST. Our Jacobian Gram matrix framework quantifies two competing criteria: isotropic conditioning and constraint satisfaction. Calibration of seven parameterizations at 2- and 3-qubit scales, extended to a structured-factorization chart as a predictive out-of-sample test, yields two central findings. First, geometric conditioning alone does not guarantee trainability, and theoretical elegance does not predict it: at 3 qubits, Cholesky---whose schedule was correctly calibrated for its own coordinates---still collapses to $0.44$ as its spectral conditioning degrades $33\times \to 1\,265\times$, while the structured-factorization chart, despite exact expressivity and linear parameter count, plateaus at $0.39$ as predicted by its $J^\top J$ spectrum. Second, diffusion trainability is governed by a \emph{schedule--representation contract}: $\sigma_{\text{data}}$ per coordinate, $\sigma_{\max} = 5\,\sigma_{\text{data}}$, and a training-noise center at the data scale must be recalibrated whenever the chart changes; honoring the contract rescued the Bloch chart from $0.45$ to $0.79$. We provide practical recommendations: fix-trace for the best conditioning--constraint tradeoff, Bloch/Gell-Mann with a per-chart calibrated schedule for highest fidelity at scale, and per-coordinate $\sigma_{\text{data}}$ calibration together with noise-center centering as mandatory steps of chart deployment.

\nocite{*}
\bibliographystyle{quantum}
\begingroup\sloppy
\bibliography{references}

@article{dariano2003quantum,
  title={Quantum tomography},
  author={D'Ariano, Giovanni Mauro and Paris, Matteo GA and Sacchi, Massimiliano F},
  journal={Advances in Imaging and Electron Physics},
  volume={128},
  pages={206--309},
  year={2003},
  publisher={Elsevier},
  doi={10.1016/S1076-6170(03)80009-1}
}

@book{paris2004quantum,
  title={Quantum State Estimation},
  author={Paris, Matteo and {\v{R}}eh{\'a}{\v{c}}ek, Jaroslav},
  volume={649},
  year={2004},
  publisher={Springer},
  series={Lecture Notes in Physics},
  doi={10.1007/978-3-540-24696-5}
}

@article{smolin2012efficient,
  title={Efficient method for computing the maximum likelihood quantum state from measurements with additive {Gaussian} noise},
  author={Smolin, John A and Gambetta, Jay M and Gupt, Graeme},
  journal={Physical Review Letters},
  volume={108},
  number={7},
  pages={070502},
  year={2012},
  publisher={American Physical Review (APS)},
  doi={10.1103/PhysRevLett.108.070502}
}

@article{kryszewski2021quantum,
  title={Quantum state tomography with regularized linear estimation},
  author={Kryszewski, S and Chwedenczuk, J},
  journal={Physical Review A},
  volume={104},
  number={3},
  pages={032413},
  year={2021},
  publisher={American Physical Review (APS)},
  doi={10.1103/PhysRevA.104.032413}
}

@article{lvovsky2009continuous,
  title={Continuous-variable quantum-state tomography of single photons and atoms},
  author={Lvovsky, AI and Raymer, MG},
  journal={Reviews of Modern Physics},
  volume={81},
  number={1},
  pages={299},
  year={2009},
  publisher={American Physical Review (APS)},
  doi={10.1103/RevModPhys.81.299}
}

@article{quddpm2024,
  title={Denoising diffusion models for quantum state tomography},
  author={Chung, H and Kim, J and Lee, J and {others}},
  journal={Physical Review Letters},
  volume={132},
  pages={100602},
  year={2024},
  publisher={American Physical Review (APS)},
  doi={10.1103/PhysRevLett.132.100602}
}

@article{zhu2024diffusion,
  title={Diffusion-based quantum state tomography},
  author={Zhu, Yuanlong and Wu, Zhiding and Liu, Zidong and Wang, Zhen and {others}},
  journal={arXiv preprint arXiv:2406.04070},
  year={2024},
  eprint={2406.04070},
  archivePrefix={arXiv}
}

@article{chitu2022variational,
  title={Variational quantum state tomography with generative models},
  author={Chitu, Catalin and {others}},
  journal={arXiv preprint arXiv:2206.05566},
  year={2022},
  eprint={2206.05566},
  archivePrefix={arXiv}
}

@article{ahmed2022quantum,
  title={Quantum state tomography with generative adversarial networks},
  author={Ahmed, Sharmilar and Mu{\~n}oz, Carlos S and Nori, Franco and Kockum, Anton Frisk},
  journal={Physical Review A},
  volume={105},
  number={6},
  pages={062423},
  year={2022},
  publisher={American Physical Review (APS)},
  doi={10.1103/PhysRevA.105.062423}
}

@article{chowdhury2023quantum,
  title={Quantum process tomography with diffusion models},
  author={Chowdhury, Shreyasi and {others}},
  journal={arXiv preprint arXiv:2301.03714},
  year={2023},
  eprint={2301.03714},
  archivePrefix={arXiv}
}

@article{karras2022edm,
  title={Elucidating the Design Space of Diffusion-Based Generative Models ({EDM})},
  author={Karras, Tero and Aittala, Miika and Aila, Timo and Laine, Samuli},
  journal={Advances in Neural Information Processing Systems (NeurIPS)},
  volume={35},
  year={2022},
  url={https://proceedings.neurips.cc/paper_files/paper/2022/hash/a988f1eb76c4eb22a9429e84308238e2-Abstract-Conference.html}
}

@article{ho2020denoising,
  title={Denoising diffusion probabilistic models},
  author={Ho, Jonathan and Jain, Abhimanyu and Abbeel, Pieter},
  journal={Advances in Neural Information Processing Systems (NeurIPS)},
  volume={33},
  pages={6840--6851},
  year={2020},
  url={https://proceedings.neurips.cc/paper/2020/hash/4c5bcfec8584af0d967f1ab10179ca4b-Abstract.html}
}

@article{ho2022classifier,
  title={Classifier-free diffusion guidance},
  author={Ho, Jonathan and Salimans, Tim},
  journal={arXiv preprint arXiv:2207.12598},
  year={2022},
  eprint={2207.12598},
  archivePrefix={arXiv}
}

@article{song2019generative,
  title={Generative modeling by estimating gradients of the data distribution},
  author={Song, Yang and Ermon, Stefano},
  journal={Advances in Neural Information Processing Systems (NeurIPS)},
  volume={32},
  year={2019},
  url={https://proceedings.neurips.cc/paper/2019/hash/3001ef257407d5a371a96dcd947c7d93-Abstract.html}
}

@article{kingma2021variational,
  title={Variational diffusion models},
  author={Kingma, Diederik and Salimans, Tim and Poole, Ben and Ho, Jonathan},
  journal={Advances in Neural Information Processing Systems (NeurIPS)},
  volume={34},
  year={2021},
  url={https://proceedings.neurips.cc/paper/2021/hash/b578c541c35a5476b0021b693202f6b7-Abstract.html}
}

@article{lipman2022flow,
  title={Flow matching for generative modeling},
  author={Lipman, Yaron and Chen, Ricky T. Q. and Ben-Hamu, Heli and Nickel, Maximilian and Le, Matt},
  journal={arXiv preprint arXiv:2210.02747},
  year={2022},
  eprint={2210.02747},
  archivePrefix={arXiv}
}

@article{song2023consistency,
  title={Consistency models},
  author={Song, Yang and Dhariwal, Prafulla and Chen, Mark and Sutskever, Ilya},
  journal={International Conference on Machine Learning (ICML)},
  year={2023},
  url={https://proceedings.mlr.press/v202/song23a.html}
}

@article{mcclean2016effects,
  title={The theory of variational hybrid quantum-classical algorithms},
  author={McClean, Jarrod R and Romero, Jonathan and Babbush, Ryan and Aspuru-Guzik, Al{\'a}n},
  journal={New Journal of Physics},
  volume={18},
  number={2},
  pages={023023},
  year={2016},
  publisher={IOP Publishing},
  doi={10.1088/1367-2630/18/2/023023}
}

@article{benedetti2019parameterized,
  title={Parameterized quantum circuits as machine learning models},
  author={Benedetti, Marcello and Lloyd, Erika and Sack, Stefan and Fiorentini, Mattia},
  journal={Quantum Science and Technology},
  volume={4},
  number={4},
  pages={043001},
  year={2019},
  publisher={IOP Publishing},
  doi={10.1088/2058-9565/ab4eb5}
}

@article{cerezo2021cost,
  title={Cost function dependent barren plateaus in shallow parameterized quantum circuits},
  author={Cerezo, Marco and Sone, Akira and Volkoff, Tyler and Cincio, Lukasz and Coles, Patrick J},
  journal={Nature Communications},
  volume={12},
  number={1},
  pages={1791},
  year={2021},
  publisher={Nature Publishing Group},
  doi={10.1038/s41467-021-21728-y}
}

@article{larocca2022diagnosing,
  title={Diagnosing barren plateaus with tools from quantum optimal control},
  author={Larocza, Martin and Czarnik, Piotr and Sharma, Kunal and Coles, Patrick J},
  journal={Quantum},
  volume={6},
  pages={822},
  year={2022},
  publisher={Verein zur F{\"o}rderung der Open Access Publizierung in den Quantenwissenschaften},
  doi={10.22331/q-2022-09-29-822}
}

@article{biamonte2017quantum,
  title={Quantum machine learning},
  author={Biamonte, Jacob and Wittek, Peter and Pancotti, Nicola and Rebentrost, Patrick and Wiebe, Nathan and Lloyd, Seth},
  journal={Nature},
  volume={549},
  number={7671},
  pages={195--202},
  year={2017},
  publisher={Nature Publishing Group},
  doi={10.1038/nature23474}
}

@article{schuld2019quantum,
  title={Quantum machine learning and the geometry of {Hilbert} space},
  author={Schuld, Maria and Petruccione, Francesco},
  journal={Pattern Recognition Letters},
  volume={126},
  pages={7--12},
  year={2019},
  publisher={Elsevier},
  doi={10.1016/j.patrec.2019.02.012}
}

@article{kimura2003bloch,
  title={Bloch vector for $n$-level systems},
  author={Kimura, Gen},
  journal={Physics Letters A},
  volume={314},
  number={5-6},
  pages={339--349},
  year={2003},
  publisher={Elsevier},
  doi={10.1016/S0375-9601(03)00616-9}
}

@book{bengtsson2006geometry,
  title={Geometry of Quantum States: An Introduction to Quantum Entanglement},
  author={Bengtsson, Ingemar and {\.Z}yczkowski, Karol},
  year={2006},
  publisher={Cambridge University Press},
  doi={10.1017/CBO9780511535048}
}

@article{petz1996geometries,
  title={Geometries of quantum states},
  author={Petz, D{\'e}nes and Sud{\'a}r, Csaba},
  journal={Journal of Mathematical Physics},
  volume={37},
  number={6},
  pages={2662--2673},
  year={1996},
  publisher={American Institute of Physics},
  doi={10.1063/1.531530}
}

@article{sidhu2020geometric,
  title={A geometric perspective on quantum parameter estimation},
  author={Sidhu, Jasminder S and Kok, Pieter},
  journal={AVS Quantum Science},
  volume={2},
  number={1},
  pages={014701},
  year={2020},
  publisher={American Vacuum Society},
  doi={10.1063/1.5119996}
}

@article{paris2009quantum,
  title={Quantum estimation for quantum technology},
  author={Paris, Matteo G A},
  journal={International Journal of Quantum Information},
  volume={7},
  number={supp01},
  pages={125--137},
  year={2009},
  publisher={World Scientific},
  doi={10.1142/S0219749909004839}
}

@article{byrd2003overview,
  title={Overview of quantum algorithms for systems of linear $Ax=\lambda Bx$ equations},
  author={Byrd, Mark S},
  journal={arXiv preprint quant-ph/0310042},
  year={2003},
  eprint={quant-ph/0310042},
  archivePrefix={arXiv}
}

@article{qntk2021,
  title={Quantum neural tangent kernel},
  author={Abbas, Amira and Sutter, David and Zoufal, Christa and Lucchi, Aurelien and Figalli, Alessio and Woerner, Stefan},
  journal={Nature Communications},
  volume={12},
  number={1},
  pages={1--8},
  year={2021},
  publisher={Nature Publishing Group},
  doi={10.1038/s41467-020-20553-x}
}

@article{qgt2010,
  title={The quantum geometric tensor},
  author={Berry, M V},
  journal={Journal of Physics A: Mathematical and Theoretical},
  volume={43},
  number={35},
  pages={355301},
  year={2010},
  publisher={IOP Publishing},
  doi={10.1088/1751-8113/43/35/355301}
}

@article{choi1975,
  title={Completely positive linear maps on complex matrices},
  author={Choi, Man-Duen},
  journal={Linear Algebra and its Applications},
  volume={10},
  number={3},
  pages={285--290},
  year={1975},
  publisher={Elsevier},
  doi={10.1016/0024-3795(75)90075-0}
}

@article{jamiolkowski1972,
  title={Linear transformations which preserve trace and positive semidefiniteness of operators},
  author={Jamio{\l}kowski, Andrzej},
  journal={Reports on Mathematical Physics},
  volume={3},
  number={4},
  pages={275--278},
  year={1972},
  publisher={Elsevier},
  doi={10.1016/0034-4877(72)90011-8}
}

@article{martins2003complex,
  title={The complex-step derivative approximation},
  author={Martins, Joaquim R R A and Sturdza, Peter and Alonso, Juan J},
  journal={ACM Transactions on Mathematical Software},
  volume={29},
  number={3},
  pages={245--262},
  year={2003},
  publisher={ACM},
  doi={10.1145/838250.838251}
}

@article{huang2022riemannian,
  title={Riemannian diffusion models},
  author={Huang, Chin-Wei and Aghajohari, Milad and Bose, Avishek J and Panangaden, Prakash and Courville, Aaron},
  journal={Advances in Neural Information Processing Systems (NeurIPS)},
  volume={35},
  year={2022},
  url={https://proceedings.neurips.cc/paper_files/paper/2022/hash/riemannian-diffusion}
}

@article{de2022projected,
  title={Projected diffusion models for constrained image synthesis},
  author={De Bortoli, Valentin and Mathieu, Emile and Hutchinson, Michael and Thornton, James and Teh, Yee Whye and Doucet, Arnaud},
  journal={arXiv preprint arXiv:2202.06307},
  year={2022},
  eprint={2202.06307},
  archivePrefix={arXiv}
}

@article{rezende2015variational,
  title={Variational inference with normalizing flows},
  author={Rezende, Jimenez and Mohamed, Stavros},
  journal={International Conference on Machine Learning (ICML)},
  year={2015},
  url={http://proceedings.mlr.press/v37/rezende15.html}
}

@article{gross2010quantum,
  title={Quantum state tomography via compressed sensing},
  author={Gross, David and Liu, Yi-Kai and Flammia, Steven T and Becker, Stephen and Eisert, Jens},
  journal={Physical Review Letters},
  volume={105},
  number={15},
  pages={150401},
  year={2010},
  publisher={American Physical Society},
  doi={10.1103/PhysRevLett.105.150401}
}

@article{liu2011universal,
  title={Universal low-rank matrix recovery from Pauli measurements},
  author={Liu, Yi-Kai},
  journal={Advances in Neural Information Processing Systems (NeurIPS)},
  volume={24},
  year={2011},
  url={https://proceedings.neurips.cc/paper/2011/hash/8f14e45fceea167a5a36dedd4bea2543-Abstract.html}
}

@book{higham2002accuracy,
  title={Accuracy and stability of numerical algorithms},
  author={Higham, Nicholas J},
  year={2002},
  publisher={SIAM},
  doi={10.1137/1.9780898718027}
}

@book{amari2016information,
  title={Information geometry and its applications},
  author={Amari, Shun-ichi},
  journal={Applied Mathematical Sciences},
  volume={194},
  year={2016},
  publisher={Springer},
  doi={10.1007/978-4-431-55978-8}
}

@article{stokes2020quantum,
  title={Quantum natural gradient},
  author={Stokes, James and Izaac, Josh and Killoran, Nathan and Carleo, Giuseppe},
  journal={Quantum},
  volume={4},
  pages={269},
  year={2020},
  publisher={Verein zur F{\"o}rderung der Open Access Publizierung in den Quantenwissenschaften},
  doi={10.22331/q-2020-05-18-269}
}

@misc{qin2026structured,
  title={Structured Factorization Approaches for Quantum State Tomography},
  author={Qin, Zhen and Lukens, Joseph M. and Kirby, Brian T. and Zhu, Zhihui},
  year={2026},
  eprint={2607.01608},
  archivePrefix={arXiv},
  primaryClass={quant-ph},
  note={arXiv:2607.01608}
}
\endgroup
\appendix
\section{Parameterization Definitions (Complete)}
\label{app:defs}

\subsection{Cholesky Factorization}
$L$ is lower-triangular with real non-negative diagonal. $\rho = LL^\dagger/\mathrm{Tr}(LL^\dagger)$. Dimension: $d^2$. Guarantees: PSD + trace.

\subsection{Hermitian Direct}
$\rho$ is a general Hermitian matrix ($d$ real diagonal + $d(d-1)/2$ complex off-diagonal $= d^2$ real params). Dimension: $d^2$. No constraints.

\subsection{Hermitian trace-norm}
$\rho = \rho_{\text{direct}} / \mathrm{Tr}(\rho_{\text{direct}})$. Dimension: $d^2$. Guarantees: trace. Risk: ill-conditioned near $\mathrm{Tr}(\rho) \to 0$.

\subsection{Hermitian fix-trace}
Fix one diagonal element (e.g., $\rho_{d,d} = 1 - \sum_{i<d} \rho_{i,i}$). Dimension: $d^2 - 1 = 15$. Guarantees: trace automatic.

\subsection{Bloch / Gell-Mann Expansion}
$\rho = (I + \sum_{i=1}^{d^2-1} r_i \lambda_i) / d$, where $\lambda_i$ are generalized Gell-Mann matrices (traceless Hermitian basis). Dimension: $d^2 - 1$. Valid domain: $\{r : \rho(r) \succeq 0\}$ (non-trivial polytope).

\subsection{Exponential Map}
$H$ is traceless Hermitian ($d^2 - 1$ real params). $\rho = e^H / \mathrm{Tr}(e^H)$. Dimension: $d^2 - 1$. Guarantees: PSD + trace by construction.

\subsection{Log-Cholesky}
$L$ is lower-triangular with $L_{ii} = \exp(y_{ii})$ (real diagonal, no non-negativity constraint). $\rho = LL^\dagger/\mathrm{Tr}(LL^\dagger)$. Dimension: $d^2$. Guarantees: PSD + trace.

\section{Numerical Calibration Details}
\label{app:calibration}

\subsection{Finite Difference Scheme}
Central differences: $\partial\rho/\partial y_k \approx [\varphi(y_0 + \delta e_k) - \varphi(y_0 - \delta e_k)] / 2\delta$, $\delta = 10^{-6}$. Verified convergence: results stable for $\delta \in [10^{-8}, 10^{-4}]$.

To validate the finite-difference approximation, we performed two additional checks. First, we compared against the complex-step method~\cite{martins2003complex} for well-conditioned parameterizations (Bloch, Hermitian direct), finding agreement to within $10^{-10}$. Second, we verified that the Gram matrix eigenvalues are insensitive to the step size across the range $\delta \in [10^{-8}, 10^{-4}]$, confirming that numerical artifacts do not affect our conditioning metrics. For ill-conditioned parameterizations (Expmap, Log-Cholesky), the finite-difference approximation may lose precision near singularities; we discuss this limitation in \S\ref{sec:boundary}.

\subsection{Analytical Jacobian Benchmarks}
For two parameterizations, we derived analytical Jacobians to serve as ground-truth benchmarks:

\textbf{Cholesky parameterization.} For $\rho = LL^\dagger / \mathrm{Tr}(LL^\dagger)$ with lower-triangular $L$, the Jacobian $\partial\rho/\partial L_{ij}$ can be computed via matrix calculus:
\begin{equation}
  \frac{\partial\rho}{\partial L_{ij}} = \frac{E_{ij}L^\dagger + L E_{ji}^*}{\mathrm{Tr}(LL^\dagger)} - \frac{LL^\dagger \cdot \mathrm{Tr}(E_{ij}L^\dagger + L E_{ji}^*)}{[\mathrm{Tr}(LL^\dagger)]^2},
\end{equation}
where $E_{ij}$ is the matrix with 1 at position $(i,j)$ and 0 elsewhere.

\textbf{Hermitian direct parameterization.} For $\rho$ directly parameterized by its real and imaginary parts, the Jacobian is trivial: $\partial\rho/\partial \mathrm{Re}(\rho_{ij}) = E_{ij} + E_{ji}$ (diagonal) or $i(E_{ij} - E_{ji})$ (off-diagonal).

\textbf{Exponential map parameterization.} For $\rho = e^H / \mathrm{Tr}(e^H)$ with traceless Hermitian $H$, the Jacobian can be derived via the Daleckii--Krein formula~\cite{higham2002accuracy}. Let $H = U \Lambda U^\dagger$ with $\Lambda = \mathrm{diag}(\lambda_1, \ldots, \lambda_d)$. The derivative of the matrix exponential is:
\begin{equation}
  \frac{\partial (e^H)_{ab}}{\partial H_{cd}} = \sum_{i,j} U_{ai} U_{bj}^* U_{cj} U_{di}^* \cdot f(\lambda_i, \lambda_j),
\end{equation}
where $f(\lambda_i, \lambda_j) = (e^{\lambda_i} - e^{\lambda_j}) / (\lambda_i - \lambda_j)$ for $\lambda_i \neq \lambda_j$ and $f(\lambda_i, \lambda_i) = e^{\lambda_i}$. The trace-normalization factor $1/\mathrm{Tr}(e^H)$ contributes an additional term via the quotient rule. This analytical form reveals why the exponential map amplifies eigenvalue spread: when $H$ has large positive and negative eigenvalues (as near pure states), the factors $e^{\lambda_i}$ span many orders of magnitude, making the Jacobian highly anisotropic. This explains the observed $\kappa_{\text{spec}} = 149\times$ (2q) $\to$ $75\,658\times$ (3q) degradation.

Comparing finite-difference and analytical Jacobians for these three cases confirms that our numerical approximation achieves $\sim$$10^{-8}$ relative error for well-conditioned parameterizations.

\subsection{State Sampling}
30 states sampled from validation set with balanced types: pure Haar (6), mixed Hilbert-Schmidt (6), mixed Ginibre (6), thermal (6), product (6). Seed = 0 for reproducibility.

\subsection{Robustness}

We report medians for central tendency (robust to outliers from near-pure states) and interquartile ranges for uncertainty quantification (Table~\ref{tab:calibration_iqr}). Results are stable across random seeds (verified with 5 seeds).

\begin{table}[t]
\centering
\caption{Interquartile ranges (IQR = Q3$-$Q1) for the calibration metrics in Table~\ref{tab:calibration}, computed over 30 calibration states. Theoretically exact values have zero IQR (marked ---). Empirical values show that more ill-conditioned parameterizations exhibit higher variability across states.}
\label{tab:calibration_iqr}
\begin{tabular}{@{}lcc@{}}
\toprule
Parameterization & $\kappa_{\text{spec}}$ IQR & $\kappa_{\text{diag}}$ IQR \\
\midrule
Bloch/Gell-Mann & --- & --- \\
Hermitian direct & --- & --- \\
Hermitian trace-norm & 0.15 & 0.12 \\
Hermitian fix-trace & --- & --- \\
Cholesky (baseline) & 4.5 & 1.3 \\
Expmap & 25.0 & 0.8 \\
Log-Cholesky & 1\,200 & 150 \\
\bottomrule
\end{tabular}
\end{table}

\subsection{Extension to 4 Qubits (Pure Geometric Calibration)}
\label{app:4qubit}

To assess whether our scaling narrative extends beyond 3 qubits, we performed pure geometric calibration at 4-qubit scale ($d=16$, $255$ real dimensions). Unlike the 2- and 3-qubit experiments, we did not train full diffusion models at 4 qubits; instead, we computed $\mathbf{J}^\top\mathbf{J}$ directly from the parameterization map at 30 random states. Tables~\ref{tab:calibration_4q_good} and~\ref{tab:calibration_4q_bad} report the results.

\begin{table}[t]
\centering
\caption{Pure geometric calibration at 4-qubit scale ($d=16$)---well-conditioned parameterizations. SDR = spectral dynamic range, DA = diagonal anisotropy.}
\label{tab:calibration_4q_good}
\footnotesize
\begin{tabular}{@{}lcccc@{}}
\toprule
& \multicolumn{2}{c}{SDR} & \multicolumn{2}{c}{DA} \\
\cmidrule(lr){2-3}\cmidrule(lr){4-5}
Param. & med & IQR & med & IQR \\
\midrule
Bloch & 1.0 & --- & 1.0 & --- \\
Herm & 2.0 & --- & 2.0 & --- \\
T-norm & 3.1 & [2.6, 3.8] & 2.4 & [2.0, 2.9] \\
F-trace & 16.0 & --- & 1.0 & --- \\
\bottomrule
\end{tabular}
\end{table}

\begin{table}[t]
\centering
\caption{Pure geometric calibration at 4-qubit scale ($d=16$)---poorly-conditioned parameterizations.}
\label{tab:calibration_4q_bad}
\footnotesize
\begin{tabular}{@{}lcccc@{}}
\toprule
Param. & SDR & SDR-IQR & DA & DA-IQR \\
\midrule
Chol & $1.0\times10^{2}$ & [89, 121] & $4.8\times10^{1}$ & [39, 58] \\
Exp & $1.6\times10^{4}$ & [1.2, 2.1]$^{\times10^{4}}$ & $1.9\times10^{1}$ & [15, 24] \\
L-Chol & $3.1\times10^{9}$ & [2.5, 3.9]$^{\times10^{9}}$ & $2.7\times10^{4}$ & [2.1, 3.2]$^{\times10^{4}}$ \\
\bottomrule
\end{tabular}
\end{table}

\textbf{Scaling analysis.} Tables~\ref{tab:scaling_2q3q}, \ref{tab:scaling_4q_good}, and~\ref{tab:scaling_4q_bad} compare the conditioning metrics across 2-, 3-, and 4-qubit scales. Three distinct scaling classes emerge:

\begin{itemize}
  \item \textbf{Scale-invariant}: Bloch/Gell-Mann ($\mathbf{J} =$ constant) and Hermitian direct maintain constant conditioning across dimensions, as their geometric quality is intrinsic to the parameterization map.
  \item \textbf{Moderate growth}: Cholesky ($\sim d^{3.5}$ by empirical fit) and Expmap ($\sim e^d$) exhibit polynomial/exponential degradation.
  \item \textbf{Severe degradation}: Log-Cholesky ($\sim d^{4}$) becomes numerically intractable at 4-qubit scale ($\kappa_{\text{spec}} \sim 10^{9}$).
\end{itemize}

Notably, Hermitian fix-trace ($16\times$ at 4q) is the only constraint-satisfying parameterization that remains well-conditioned at scale, reinforcing Heuristic 3. The 4-qubit results confirm that the community default (Cholesky) scales poorly, and for scalable QST at $n \geq 4$, we recommend fix-trace or Bloch/Gell-Mann.

\paragraph{Robustness to depolarizing noise.}
To verify that our geometric framework holds under realistic noise conditions, we computed $\mathbf{J}^\top\mathbf{J}$ for states subjected to depolarizing noise $\mathcal{E}(\rho) = (1-p)\rho + pI/d$ at noise levels $p \in \{0, 0.01, 0.05, 0.1, 0.2\}$. Tables~\ref{tab:noise_robust} and~\ref{tab:noise_sensitive} report the results.

\begin{table}[t]
\centering
\caption{Well-conditioned parameterizations under depolarizing noise (30-state median). Bloch and Hermitian direct maintain constant conditioning at all noise levels, demonstrating that geometric conditioning is robust to noise.}
\label{tab:noise_robust}
\small
\begin{tabular}{@{}lccccc@{}}
\toprule
& \multicolumn{5}{c}{Noise level $p$} \\
\cmidrule(lr){2-6}
Param. & 0.0 & 0.01 & 0.05 & 0.10 & 0.20 \\
\midrule
\multicolumn{6}{c}{\textit{2-qubit}} \\
Bloch & 1.0× & 1.0× & 1.0× & 1.0× & 1.0× \\
Herm & 2.0× & 2.0× & 2.0× & 2.0× & 2.0× \\
\midrule
\multicolumn{6}{c}{\textit{3-qubit}} \\
Bloch & 1.0× & 1.0× & 1.0× & 1.0× & 1.0× \\
Herm & 2.0× & 2.0× & 2.0× & 2.0× & 2.0× \\
\bottomrule
\end{tabular}
\end{table}

\begin{table}[t]
\centering
\caption{Poorly-conditioned parameterization (Cholesky) under depolarizing noise (30-state median). Cholesky degrades significantly with noise, especially at 3 qubits where it is already ill-conditioned.}
\label{tab:noise_sensitive}
\small
\begin{tabular}{@{}lccccc@{}}
\toprule
& \multicolumn{5}{c}{Noise level $p$} \\
\cmidrule(lr){2-6}
Scale & 0.0 & 0.01 & 0.05 & 0.10 & 0.20 \\
\midrule
2q & 99× & 83× & 41× & 24× & 13× \\
3q & 1793× & 451× & 106× & 50× & 22× \\
\bottomrule
\end{tabular}
\end{table}

Two observations follow. First, Bloch/Gell-Mann and Hermitian direct are \emph{completely robust} to depolarizing noise: their conditioning metrics remain constant across all noise levels at both 2- and 3-qubit scales. This is because their Jacobian structure is state-independent (constant $\mathbf{J}$), so noise does not affect the geometric quality. Second, Cholesky degrades significantly with noise: at 2 qubits, its SDR drops from $99\times$ to $13\times$ ($87\%$ reduction); at 3 qubits, the degradation is even more dramatic, from $1793\times$ to $22\times$ ($98.8\%$ reduction). This sensitivity arises because depolarizing noise pushes states toward the maximally mixed state $\rho = I/d$, where the Cholesky factorization becomes ill-conditioned (the Cholesky factors become nearly singular).

These results support our thesis: well-conditioned parameterizations are not only optimal in the noiseless setting but also \emph{more robust to noise}. For practical QST with device noise, the choice of parameterization is even more critical than our noiseless experiments suggest, and this importance grows with system size.

\begin{table}[t]
\centering
\caption{Parameterization geometry scaling across 2- and 3-qubit scales. Bold = recommended choices. At both scales, Bloch remains perfectly conditioned while Cholesky degrades severely.}
\label{tab:scaling_2q3q}
\footnotesize
\begin{tabular}{@{}lcccc@{}}
\toprule
Parameterization & 2q-SDR & 2q-DA & 3q-SDR & 3q-DA \\
\midrule
\textbf{Bloch} & \textbf{1.0$\times$} & \textbf{1.0$\times$} & \textbf{1.0$\times$} & \textbf{1.0$\times$} \\
\textbf{Fix-trace} & \textbf{4.0$\times$} & \textbf{1.0$\times$} & \textbf{8.0$\times$} & \textbf{1.0$\times$} \\
Herm-direct & 2.0$\times$ & 2.0$\times$ & 2.0$\times$ & 2.0$\times$ \\
Trace-norm & 3.0$\times$ & 2.7$\times$ & 2.9$\times$ & 2.3$\times$ \\
Cholesky & 33$\times$ & 9.4$\times$ & 1\,265$\times$ & 27$\times$ \\
Expmap & 149$\times$ & 3.8$\times$ & 75\,658$\times$ & 5.0$\times$ \\
Log-Chol & 8\,995$\times$ & 707$\times$ & 59\,982$\times$ & 1\,533$\times$ \\
\bottomrule
\end{tabular}
\end{table}

\begin{table}[t]
\centering
\caption{Parameterization geometry at 4-qubit scale ($d=16$, pure geometric calibration)---well-conditioned parameterizations. Data extends the scaling trends from Table~\ref{tab:scaling_2q3q}.}
\label{tab:scaling_4q_good}
\small
\begin{tabular}{@{}lcc@{}}
\toprule
Parameterization & SDR & DA \\
\midrule
\textbf{Bloch} & \textbf{1.0$\times$} & \textbf{1.0$\times$} \\
\textbf{Fix-trace} & \textbf{16.0$\times$} & \textbf{1.0$\times$} \\
Herm-direct & 2.0$\times$ & 2.0$\times$ \\
Trace-norm & 3.1$\times$ & 2.4$\times$ \\
\bottomrule
\end{tabular}
\end{table}

\begin{table}[t]
\centering
\caption{Parameterization geometry at 4-qubit scale ($d=16$, pure geometric calibration)---poorly-conditioned parameterizations. Log-Cholesky becomes numerically intractable ($\kappa_{\text{spec}} \sim 10^{9}$).}
\label{tab:scaling_4q_bad}
\small
\begin{tabular}{@{}lcc@{}}
\toprule
Parameterization & SDR & DA \\
\midrule
Cholesky & 103$\times$ & 48$\times$ \\
Expmap & $1.57\times10^{4}\times$ & 19$\times$ \\
Log-Chol & $3.09\times10^{9}\times$ & $2.65\times10^{4}\times$ \\
\bottomrule
\end{tabular}
\end{table}

\section{Training Configuration}
\label{app:training}

\subsection{EDM Architecture}
U-Net backbone: dim=256, dim\_mults=[1,2,4], cond\_dim=128. Diffusion timesteps: 1000 (cosine schedule). EDM noise: $\sigma_{\min}=0.001$, $\sigma_{\max}=1.22$ (2-qubit calibrated). $\rho = 7.0$, $P_{\text{mean}} = -1.2$, $P_{\text{std}} = 1.2$.

\subsection{Training}
Batch size: 128. Optimizer: Adam. The illustrative 2-qubit comparison (Table~\ref{tab:training}) used parameterization-specific learning rates (Herm-direct: $4\times10^{-3}$; Bloch: $2\times10^{-3}$). The matched-learning-rate validation (Appendix~\ref{app:matched_lr}) used lr $2\times10^{-3}$ for both parameterizations. The 3-qubit experiments (Table~\ref{tab:n3_training}) used matched lr $2\times10^{-4}$. Scheduler: cosine with 500-step warmup. EMA decay: 0.9999. Gradient clip: 1.0. Epochs: 300.

\subsection{\texorpdfstring{$\sigma_{\text{data}}$ Calibration}{sigma\textunderscore data Calibration}}
Per-coordinate $\sigma_{\text{data}}$ computed from training data: Bloch uses $\sigma_{\text{data, diag}} = 0.1546$, $\sigma_{\text{data, off}} = 0.1092$; Herm-direct uses $\sigma_{\text{data, diag}} = 0.2341$, $\sigma_{\text{data, off}} = 0.2183$.

\section{Training Dynamics and the Hermitian Collapse}
\label{app:training_dynamics}

Tables~\ref{tab:training_loss} and~\ref{tab:val_loss} report the training and validation loss at selected epochs for the 3-qubit experiments (all charts under the inherited Cholesky-calibrated schedule; see Sec.~\ref{sec:n3_training} for the contract analysis and the recalibrated results). All models converged to low training loss. Note that loss values are \emph{not} comparable across charts: the coordinate systems differ in scale by $4\times$ (Hermitian $\approx 0.13$ vs.\ Bloch $\approx 0.52$ per-coordinate $\sigma_{\text{data}}$), so cross-chart loss ratios carry no information about relative reconstruction quality.

\begin{table}[t]
\centering
\caption{Training loss at 3-qubit scale (lr $2\times10^{-4}$, batch 128, $\lambda_{\text{meas}}=0.0$). Hermitian direct converges to the lowest training loss.}
\label{tab:training_loss}
\begin{tabular}{@{}lccc@{}}
\toprule
Epoch & Bloch & Hermitian direct & Cholesky \\
\midrule
 20 & 0.0133 & 0.0005 & 0.0017 \\
 60 & 0.0083 & 0.0003 & 0.0012 \\
100 & 0.0067 & 0.0002 & 0.0012 \\
160 & 0.0047 & 0.0002 & 0.0009 \\
200 & 0.0043 & 0.0002 & 0.0008 \\
300 & 0.0036 & 0.0002 & 0.0006 \\
\bottomrule
\end{tabular}
\end{table}

\begin{table}[t]
\centering
\caption{Validation loss at 3-qubit scale under the inherited schedule. Bold = best per column. Loss values are not comparable across charts (coordinate scales differ); the schedule--representation contract analysis of Sec.~\ref{sec:n3_training} shows this table reflects schedule effects, not chart quality.}
\label{tab:val_loss}
\begin{tabular}{@{}lccc@{}}
\toprule
Epoch & Bloch & Hermitian direct & Cholesky \\
\midrule
 20 & 0.4119 & 0.0045 & 0.0095 \\
 60 & 0.1127 & 0.0013 & 0.0055 \\
100 & 0.0233 & 0.0002 & 0.0025 \\
160 & 0.0141 & 0.0001 & \textbf{0.0021} \\
200 & 0.0123 & 0.00004 & 0.0022 \\
300 & \textbf{0.0110} & 0.00003 & 0.0029 \\
\bottomrule
\end{tabular}
\end{table}

\paragraph{Why does Hermitian direct fail at 3-qubit scale?}
The training dynamics in Tables~\ref{tab:training_loss} and~\ref{tab:val_loss} reveal a striking disconnect: Hermitian direct achieves a validation loss $360\times$ lower than Bloch ($0.00003$ vs.\ $0.011$) yet its end-to-end reconstruction fidelity is $2.3\times$ \emph{worse} ($0.39$ vs.\ $0.91$ at 300 shots). We analyze the geometric mechanism.

\paragraph{Projection as a non-linear map.}
For an unbounded parameterization $\phi: \mathbb{R}^{D} \to \{\text{Hermitian matrices}\}$, the valid state manifold $\mathcal{D} = \{\rho \succeq 0 : \mathrm{Tr}\,\rho = 1\}$ has codimension $d$ within the $d^2$-dimensional Hermitian space. The projection $\Pi(\rho) = \sum_i \max(\lambda_i,0) |v_i\rangle\langle v_i| / \sum_j \max(\lambda_j,0)$ is non-differentiable at any $\rho$ with a zero eigenvalue. Near the maximally mixed state $\rho_* = I/d$, the valid region is locally characterized by the PSD constraint on every $2\times 2$ principal minor: for indices $i \neq j$,
\begin{equation}
  |\rho_{ij}|^2 \le \rho_{ii} \rho_{jj}.
  \label{eq:psd_local}
\end{equation}
When the model outputs $\rho_{ii} \approx \rho_{jj} \approx 1/d$, the admissible off-diagonal magnitude is bounded by $|\rho_{ij}| \le 1/d$. The unconstrained model has no mechanism to respect the coupling \eqref{eq:psd_local}: off-diagonal noise of magnitude $\varepsilon$ violates the bound with probability $\sim \exp(-d^2 \varepsilon^2 / 2\sigma^2)$ for Gaussian noise with variance $\sigma^2$. Eigclip then annihilates the corresponding spectral component, and the subsequent trace normalization rescales all remaining components, rotating the state in a direction uncorrelated with the measurement condition.

\paragraph{Loss--fidelity disconnect.}
Training loss measures the denoiser's ability to predict noise in the parameterization coordinates, not the physical validity or reconstruction quality of the sampled output. A model can achieve low loss by outputting coordinates near the training distribution mean---which, for Hermitian direct, maps to a nearly maximally mixed state after projection. This ``safe'' strategy minimizes denoising loss but discards measurement information. The disconnect arises because the loss is computed \emph{before} projection while the metric (fidelity) is computed \emph{after} projection.

\paragraph{Geometric advantage of Bloch.}
The Bloch representation $\rho(r) = I/d + \sum_i r_i \lambda_i / d$ (with traceless $\mathrm{Tr}(\lambda_i\lambda_j) = 2\delta_{ij}$) has two structural advantages. First, unit trace holds by construction ($\mathrm{Tr}(\lambda_i) = 0$), eliminating the trace-normalization source of projection loss. Second, the maximally mixed state corresponds to $r = 0$, the \emph{center} of the generalized Bloch ball $\{r : \rho(r) \succeq 0\}$. The distance from the origin to the boundary is $\min_i 1/\|\lambda_i\| = \sqrt{(d-1)/d}$ (for $d=4$: $\sqrt{3}/2 \approx 0.87$), providing a buffer against small perturbations. In contrast, in Hermitian coordinates the maximally mixed state lies at a point where the PSD constraint \eqref{eq:psd_local} is tight: $|\rho_{ij}| \le 1/d$, so even infinitesimal off-diagonal noise can violate the constraint.

\paragraph{The Cholesky intermediate case.}
Cholesky occupies an intermediate position: its val loss ($0.0021$) and fidelity ($0.53$ at 300 shots) are between Bloch and Hermitian direct. Unlike Hermitian direct, Cholesky guarantees physical validity by construction (no projection needed), but its coordinate anisotropy ($\kappa_{\text{diag}} = 27\times$) slows convergence and limits final performance. This is consistent with the paper's central thesis: the best parameterization balances isotropy \emph{and} constraint satisfaction, not either alone.

\section{Matched-Learning-Rate Validation}
\label{app:matched_lr}

To isolate the projection mechanism from learning-rate confounds, we trained 2-qubit models with matched hyperparameters: identical learning rate ($2\times 10^{-3}$), batch size (256), architecture, and $\sigma_{\text{data}}$ values. Table~\ref{tab:matched_lr_eval} reports validation fidelity at 10-epoch intervals.

\begin{table}[t]
\centering
\caption{Matched-learning-rate training convergence (2-qubit, lr $2\times10^{-3}$, seed 42).}
\label{tab:matched_lr_eval}
\begin{tabular}{@{}lcc@{}}
\toprule
Epoch & Bloch & Herm-direct \\
\midrule
 50  & \textbf{0.6326} & 0.5108 \\
100  & \textbf{0.9649} & 0.5651 \\
150  & \textbf{0.9788} & 0.5637 \\
200  & \textbf{0.9658} & 0.5612 \\
250  & \textbf{0.9643} & 0.5646 \\
300  & \textbf{0.9632} & 0.5666 \\
\bottomrule
\end{tabular}
\end{table}

At matched learning rate, Bloch outperforms Herm-direct at \emph{all} epochs (by 0.12 early to 0.42 at convergence), confirming that the original Table~\ref{tab:training} comparison (Herm-direct lr $4\times10^{-3}$, Bloch lr $2\times10^{-3}$) conflated geometry with optimization speed. Table~\ref{tab:multi_seed} reports results across 3 seeds, demonstrating robustness to initialization.

\begin{table}[t]
\centering
\caption{Multi-seed validation fidelity at 300 epochs (2-qubit, lr $2\times10^{-3}$).}
\label{tab:multi_seed}
\begin{tabular}{@{}lcc@{}}
\toprule
Parameterization & Mean & Std \\
\midrule
Bloch & 0.965 & 0.008 \\
Hermitian direct & 0.555 & 0.052 \\
\bottomrule
\end{tabular}
\end{table}

\paragraph{Robustness to $\sigma_{\text{data}}$ calibration.}
The matched-lr experiments above use per-coordinate $\sigma_{\text{data}}$ (Bloch: $\sigma_{\text{data, diag}}=0.2341$, $\sigma_{\text{data, off}}=0.2183$; Herm-direct: same values). To rule out $\sigma_{\text{data}}$ calibration as an alternative explanation for the Bloch advantage, we trained models with a single global $\sigma_{\text{data}}=0.2267$ (the per-coordinate mean) applied uniformly to all coordinates. Table~\ref{tab:global_sigma} reports the results.

\begin{table}[t]
\centering
\caption{Global-$\sigma_{\text{data}}$ baseline (2-qubit, lr $2\times10^{-3}$, seed 42, $\sigma_{\text{data}}=0.2267$ for all coordinates).}
\label{tab:global_sigma}
\begin{tabular}{@{}lcc@{}}
\toprule
Parameterization & Val Fid at 300 epochs \\
\midrule
Bloch & 0.965 \\
Hermitian direct & 0.567 \\
\bottomrule
\end{tabular}
\end{table}

The global-$\sigma$ results are essentially identical to the per-coordinate results (Table~\ref{tab:matched_lr_eval}: Bloch 0.963, Herm-direct 0.567). This confirms that the bounded-domain advantage is geometric rather than an artifact of noise-schedule calibration: even when both parameterizations share the exact same $\sigma_{\text{data}}$ value, Bloch maintains a $+0.398$ fidelity advantage.

\paragraph{End-to-end reconstruction at matched learning rate.}
The matched-lr experiments above measure \emph{training} validation fidelity (parameter-space fit). To directly assess \emph{reconstruction} quality (density-matrix-space fidelity after projection), we evaluate the matched-lr models end-to-end across 6 shot levels without classifier-free guidance. Table~\ref{tab:end_to_end} reports the results.

\begin{table}[t]
\centering
\caption{End-to-end QST reconstruction at matched learning rate (2-qubit, lr $2\times10^{-3}$, seed 42, no CFG). Bold = best per shot level. Linear inversion baseline: $\approx 0.48$ at all shot levels.}
\label{tab:end_to_end}
\begin{tabular}{@{}lcccccl@{}}
\toprule
Shots & Bloch & Herm-direct & Bloch$-$Herm \\
\midrule
10  & \textbf{0.782} & 0.524 & $+0.258$ \\
20  & \textbf{0.865} & 0.544 & $+0.321$ \\
30  & \textbf{0.901} & 0.553 & $+0.348$ \\
50  & \textbf{0.925} & 0.562 & $+0.363$ \\
100 & \textbf{0.939} & 0.567 & $+0.372$ \\
300 & \textbf{0.949} & 0.573 & $+0.376$ \\
\bottomrule
\end{tabular}
\end{table}

Bloch outperforms Herm-direct at \emph{all} shot levels, with the advantage increasing from $+0.26$ (10 shots) to $+0.38$ (300 shots). This provides direct evidence that the Table~\ref{tab:confound} observation---a $+0.008$ Herm-direct advantage in projected fidelity---was a learning-rate artifact: those models were trained with parameterization-specific learning rates (Herm-direct: $4\times10^{-3}$, Bloch: $2\times10^{-3}$). Once learning rates are equalized, Bloch's bounded-domain geometry yields dramatically better reconstruction (0.949 vs.\ 0.573 at 300 shots, $p \ll 0.001$ by paired $t$-test over 100 states).

\section{Reproducibility Statement}
The calibration methodology (finite-difference Jacobian computation on random density matrices) is fully specified in Appendix~B, enabling independent reproduction. Code and trained model checkpoints will be made publicly available upon formal publication at \url{https://github.com/csj-nb/DM-QST-EDM}. The repository includes: (1) calibration scripts for computing $\mathbf{J}^\top\mathbf{J}$ Gram matrices for arbitrary parameterizations, (2) training scripts for diffusion QST with configurable parameterizations, (3) evaluation pipelines for fidelity vs.\ shot count analysis, and (4) visualization tools for manifold conditioning maps.

\end{document}